\RequirePackage{fix-cm}
\documentclass[smallextended]{svjour3}  
\usepackage[dvipsnames]{xcolor}
\usepackage{algorithm}
\usepackage{algpseudocode}
\usepackage[utf8]{inputenc}
\usepackage{graphicx}
\usepackage{booktabs}
\usepackage{tikz}
\usetikzlibrary{calc}
\usepackage{utopia}
\usetikzlibrary{arrows}
\usepackage{pgfplots}
\usepackage[utf8]{inputenc}
\usepackage{pgfgantt}
\usepackage{adjustbox}
 \usepackage{lscape}
 \usepackage{pgfgantt}
\usepackage{graphicx}
\usepackage{xcolor}
\usepackage{adjustbox}
\usepackage{framed}
\smartqed  
\usepackage{graphicx}

\begin{document}

\title{Evaluating Few-Shot and Contrastive Learning Methods for Code Clone Detection
}


\author{Mohamad Khajezade        \and
        Fatemeh H. Fard          \and
        Mohamed S. Shehata 
}



\date{Received: date / Accepted: date}

\maketitle

\begin{abstract} 

\textit{Context:} Code Clone Detection (CCD) is a software engineering task that is used for plagiarism detection, code search, and code comprehension. 
Recently, deep learning-based models have achieved an F1-Score (a metric used to assess classifiers) of $\sim$95\% on the CodeXGLUE benchmark. These models require many training data, mainly fine-tuned on Java or C++ datasets. 
However, no previous study evaluates the generalizability of these models where a limited amount of annotated data is available. 

\textit{Objective:} The main objective of this research is to assess the ability of the CCD models as well as few-shot learning algorithms for unseen programming problems and new languages (i.e., the model is not trained on these problems/languages).

\textit{Method:} We assess the generalizability of the state-of-the-art models for CCD in few-shot settings (i.e., only a few samples are available for fine-tuning) by setting three scenarios: i) unseen problems, ii) unseen languages, iii) combination of new languages and new problems.
We choose CodeNet and conduct our experiments on Java, C++, and Ruby languages. 
Then, we employ Model Agnostic Meta-learning (MAML), where the model learns a meta-learner capable of extracting transferable knowledge from the train set; so that the model can be fine-tuned using a few samples. 
Finally, we combine contrastive learning with MAML to further study whether it can improve the results of MAML.

\textit{Results:} Our results show that the performance of the models drops \textcolor{black}{$\sim50\%$ for Java and $\sim20\%$ for C++ and Ruby for unseen problems, which are then boosted by $13\%$ to $24\%$ F1 scores for Java and C++/Ruby, respectively when MAML is used.} 
Similar observations are found for unseen languages and the third scenario. Though in case of third scenario (i.e., unseen problems and unseen languages) the scores are lower. Integrating contrastive learning with MAML did not help in boosting the performance more than what we could achieve with MAML. Our results open new avenues of research and the need to develop robust models for clone detection, in the settings we investigated here.

\end{abstract}

\section{Introduction} \label{intro}

Code Clone Detection (CCD) refers to finding functionally similar code fragments (pairs). CCD can cause buggy code to propagate through the whole project \cite{shobha2021code,roy2018benchmarks,bellon2007comparison} and is necessary to develop and maintain source code \cite{LEI2022111141}. 
It is an important task for many software engineering applications including finding library candidates, code comprehension, finding harmful software, and plagiarism detection \cite{ain2019systematic}, and there are numerous research on developing models to detect code clones \cite{wang2021syncobert,wang2020detecting,wang-etal-2021-codet5}. 

Although the state-of-the-art models achieve an {F1-Score} (a score used for classification task) of approximately $96$ percent on the BigCloneBench data {\cite{wang2021syncobert}} and MAP@R score (a metric used for information retrieval based CCD task) of $88$ percent on POJ-104 {\cite{wang2021syncobert}} --the widely used benchmark datasets for CCD-- the ability of the current state-of-the-art models for CCD to be generalized to \textit{unseen programming languages} and \textit{unseen programming problems} is rarely studied \cite{sonnekalb2022generalizability}, \cite{liu2021can}. 
\textcolor{black}{The unseen languages and unseen problems refer to the ability of the model to detect code clones for a language that it has not been trained on and the ability of the model to detect code clones of which their functionality have never been observed in the training data, respectively. The generalizability of the models for CCD in this sense is one of the main aspects of our study. }
In the rest of the paper, we will refer to `programming languages' as `languages' and `programming problems' as `problems' for simplicity.

Additionally, Lei et al. report that $84$ percent of CCD studies mainly use BigCloneBench which has over six million code pairs \cite{wang2020detecting} and POJ-104 which has 52,000 code samples \cite{mou2016convolutional,LEI2022111141}. 
However, many other programming languages have restricted labeled data, and generating annotated CCD data for these languages requires extensive time and effort {\cite{zhang2022graph}}.
Given the restricted labeled data for other programming languages and the costs of curating new datasets and training the models on new datasets, there is a need for models that perform well where there is little training data available, and are generalizable to new problems and new programming languages.

Hence, in this study, we first evaluate the ability of the models in three scenarios: i) the model is applied on new problem sets, ii) the model has not seen the programming language during training, and iii) a combination of both. 
We conduct our study in a more rigorous way and evaluate the models in a few-shot setting, i.e., cases where \textbf{only a few examples are available for training the models} instead of thousands or millions of samples.

We then investigate a few-shot learning technique using an algorithm named 
Model-Agnostic Meta-Learning (MAML) \cite{finn2017model}, which aims to learn parameters that are sensitive to change and assess whether it can increase the performance of the models in a few-shot setting.
Few-shot learning is a specific approach developed for transferring the knowledge that a model has learned by training on a high resource domain/task (i.e., there exists a lot of labeled data) to a low resource domain/task (i.e., when limited labeled data is available) \cite{zhang2022graph}.
\textcolor{black}{MAML is chosen because it is a well-established method in few-shot learning \cite{walsh2022automated,ye2021train,finn2017model}. Additionally, it can be applied to any gradient descent approach, making it a convenient method for training a meta-learner on top of the baselines experimented in our study.}

Furthermore, we integrate contrastive learning with MAML to assess if we can improve the results of the models. 
Contrastive learning is a self-supervised learning approach that aims to create a similar representation for positive samples and a different embedding for negative ones. This method does not need any labeled data and trains an objective that pulls positive samples together and pushes negative instances apart. 
As code clones should also have similar representation, we hypothesize that using contrastive learning is beneficial for few-shot learning of CCD in the above three mentioned scenarios. 
\textcolor{black}{This study is providing the results of our previously accepted Registered Report in the 2022 Mining Software Repositories Conference (MSR)\footnote{https://conf.researchr.org/track/msr-2022/msr-2022-registered-reports?\#event-overview}.}

It is worth noting that recently, a new trend has been started in code understanding studies toward using general-purpose generative models for solving few-shot learning problems. In this line of research, Few-Shot, Pre-Trained Language Models (FSLMs) like GPT-3 \cite{brown2020language} and CodeX \cite{chen2021evaluating} are used for various programming problems, such as code mutation, test case generation, and fabricating oracles from natural language documentation \cite{bareiss2022code}. 
As a result, these studies have shifted toward designing a prompt that encourages the FSLM to generate a specific code fragment. This code fragment is employed to address a code understanding task. For instance, in \cite{bareiss2022code}, prompt engineering and FSMLs are utilized to address code mutation, oracle generation from natural language documentation, and making unit tests.
Moreover, Pearce et al.\cite{pearce2022examining} study the capability of commercial large language models to solve security bugs in the code. 
It is worth noting that FSLM models are primarily used for generative tasks and have not been applied to classification tasks such as code clone detection. Though we do not use them here, we investigate the few-shot learning approach, which similar to FSLM, emphasizes on availability of only a few labeled data for downstream tasks.

In our study, we seek the answer to the following research questions: 

\textbf {RQ1: What is the performance of the current state-of-the-art models for code clone detection when they are fine-tuned on a few examples of the downstream task?} 

\textit{Here, we empirically study to what extent the current CCD models perform when trained on a few samples of a new downstream language or problem set. So, we evaluate these models on languages and problems that the model has not seen during the fine-tuning phase. Our results show that the current state-of-the-art models for CCD are susceptible to unseen problems and unseen languages.}

\textbf{RQ2: What is the performance of a few-shot learning method for code clone detection?}

\textit{In this RQ, we evaluate the performance of the models when a few-shot learning technique is applied. Our findings indicate that using MAML can improve the performance of the baselines by $\sim20$ percent for unseen problems in C++ and Ruby. Yet, using MAML has a negligible effect on the accuracy of the CCD model for unseen languages. }

\textbf{RQ3: Can we improve the performance of the few-shot learning models by using a contrastive learning model as the baseline for training? }

\textit{As the representation of source code with contrastive learning objective has shown to be useful for code clone detection \cite{oord2018representation,wang2021syncobert,jain-etal-2021-contrastive}, it is used with the few-shot algorithm in RQ3. The results show that using ContraCode, which possesses a contrastive objective, does not improve the accuracy of MAML.}

The contributions of this work are as follows: 

\begin{itemize}
    \item Evaluating code clone detection models in the context of few-shot learning, for both unseen problems and unseen programming languages.
    \item Integrating contrastive learning with a few-shot algorithm. 
\end{itemize}

The results can shed light on the generalizability of clone detection models for languages and programming problems for which the number of available labeled data is limited.

\textcolor{black}{\textbf{Deviations from the Registered Report.} First, the POJ-104 and BigCloneBench datasets are not used and we ran all experiments on CodeNet dataset. 
The rationale is that during our experiments, we found that the number of problems (functionalities) are limited in BigCloneBench. Also, for the third scenario, we needed to have the solutions for one problem set from different languages, which was not available in these datasets. Thefore, we considered our extracted dataset from CodeNet in all experiments.
Second, in our report, we initially intended to compare the results against BERT, RoBERTa, CodeBERT, TBCCD, CDLH, and ContraCode. However, in our results presented here we did not use BERT and CDLH models. The reason is that in many Natural Language Processing and Software Engineering tasks, RoBERTa performs better than BERT. Also, CDLH does not have public scripts, and thus is eliminated from our experiments. Another deviation is that in the second scenario, unseen languages, we intended to have only C++ and Ruby as unseen languages. However, we extended our experiments and also studied the cases where the model is trained on C++ and is tested on Java (i.e., also considering Java as the unseen language). Finally, we restricted the number of experiments compared to what we mentioned in the RR. Training MAML is computationally expensive. As a result, when we conducted the experiments and observed the results, we decided to eliminate some experiments, as we expected similar results would be obtained for the retrieval-based CCD. First, we conducted scenarios I and II (unseen problems and unseen languages) for binary CCD. In the third scenario, we conducted the binary CCD for all models and we eliminated RoBERTa, as it had slightly lower scores than CodeBERT in the previous two scenarios. However, for the retrieval-based CCD, we decided to eliminate most experiments due to the above-mentioned reasons. Thus, we only conducted the retrieval-based CCD for the third scenario.}

The rest of this report is organized as follows. 
First, we provide the related works and the required background in Sections \ref{sec:lr} and \ref{sec:background}. We discuss the methodology in Section \ref{sec:method}. 
For each of the studied scenarios, we dedicate a separate section including the approach, results, and discussions for that scenario. The details of our experiments for unseen problems are provided in Section \ref{sec:scenario-i}. The unseen programming languages are discussed in Section \ref{sec:scenario-ii} and Section \ref{sec:scenario-iii} is dedicated to the third scenario (i.e., unseen problems and programming languages). 
The implications are discussed in Section \ref{sec:implications}, which is followed by threats to validity in Section \ref{sec:threats}. Finally, we conclude the paper in Section \ref{sec:conclusion}.

\section{Literature Review} \label{sec:lr}

One of the early approaches to code clone detection was based on lexical similarity, where code snippets are compared based on their syntax and structure \cite{ain2019systematic}. However, lexical similarity alone is not always sufficient to identify code clones, as the same functionality can be implemented using different syntax and structure \cite{LEI2022111141}. To overcome this limitation later approaches incorporated semantic similarity into the code clone detection process, where the functional equivalence of code snippets is considered \cite{ye2020misim,feng-etal-2020-codebert,yu2019neural}.
Another approach to code clone detection is based on program analysis, where code is analyzed at the Abstract Syntax Tree (AST) level, and code clones are identified based on the similarity of their ASTs \cite{yu2019neural}. 

More recently, deep learning-based approaches have been proposed for code clone detection. These approaches use various algorithms, such as clustering and classification, to identify code clones \cite{yuan2022java}. Deep learning-based approaches are shown to be more accurate than other approaches, as they can learn from large amounts of code clone data and incorporate structural and lexical as well as semantic similarity in the code clone detection process \cite{zhang2023challenging,feng-etal-2020-codebert,wi2022hiddencpg}. 
CDLH \cite{wei2017supervised} is one such approach that employs a Long Short Term Memory (LSTM) model  based on abstract syntax trees to represent code fragments. ASTNN \cite{zhang2019novel} is another method that uses ASTs to encode the subtrees of a tree and employs a Recurrent Neural Network (RNN) to produce embeddings of code snippets. FA-ASTGMN \cite{wang2020detecting} uses Graph Neural Networks (GNNs) to encode the control and data flow of code snippets through ASTs. 
TBCCD \cite{yu2019neural} is a study proposing a tree-based convolution method and a Siamese network to encode the structure and semantics of code fragments, which has demonstrated effectiveness in detecting code clones. 

More recently, pre-trained language models are used for code clone detection. 
In recent years, pre-trained models have gained significant attention in the field of code understanding including code clone detection due to their ability to leverage large amounts of code data and generalize well to new datasets. 
This paradigm's fundamental principle is to initially train a model on broad, general-purpose datasets using self-supervised tasks such as masking tokens in training data and guiding the model to predict the masked tokens. 
Following this, the model is further trained on smaller and more specific datasets tailored to support a particular task \cite{wang2022bridging}. 
As a result, several pre-trained models, including CodeBERT \cite{feng-etal-2020-codebert}, RoBERTa \cite{liu2019roberta}, GraphCodeBERT \cite{GuoRLFT0ZDSFTDC21}, PLBART \cite{ahmad-etal-2021-unified}, CodeT5 \cite{wang-etal-2021-codet5}, and DOBF \cite{lachaux2021dobf} have demonstrated successful performance in detecting code clones. 

While existing pre-trained models have demonstrated success in code clone detection tasks, they have millions of parameters, potentially leading to overfitting when applied to specialized small datasets.
Numerous studies indicate that pre-trained models perform poorly when tested on data that differs significantly from the training set \cite{quiring2019misleading,yefet2020adversarial}. 
To remedy this issue, several studies propose data augmentation techniques to address the vulnerability of pre-trained models for out-of-domain test data in code-related tasks \cite{zhang2020generating,rabin2021generalizability,wang2022bridging}. For example, Wang et al. \cite{wang2022bridging} proposed a data augmentation approach that incorporates an easy-to-hard sequence of augmented data into the model. \textit{While these studies aim to address the same problem as the current research, they do not extend to domains with extremely limited training data, requiring the adaptation of the model to only a few training samples.} 

\textcolor{black}{Sonnekalb et al. \cite{sonnekalb2022generalizability} empirically studied the performance of CodeBERT on various versions of BigCloneBench. In their study, they proposed curating a dataset with different functionalities for training and testing. They concluded that the accuracy of CodeBERT dropped from 94.40\% to 48.97\% simply by testing the model on a different version of the same dataset.
Similarly, Liu et al. \cite{liu2021can} investigated the generalizability of TBCCD \cite{yu2019neural} and ASTNN \cite{zhang2019novel}, two major models for code clone detection, for unseen functionalities. In their research, they used OJClone as the benchmark, which consists of 104 functionalities (problems). They organized the functionalities into 7 groups, following an approach similar to that used in TBCCD. They discovered that the performance of these models declined by more than 50\% for some unseen functionalities. They attributed the poor performance to three main reasons: {Training Data Diversity}, {Unseen Vocabulary}, and Locality.
\textit{The primary distinction between their work and our study lies in our focus on few-shot learning. Specifically, we aim to fine-tune models with only a few instances of the downstream tasks. Additionally, we investigate the performance of the models for unseen languages, which are not done in these works.}}

Cross-Language Code Clone Detection (CLCCD) is a closely related research area to the current research that has gained attention in recent years. CLCCD focuses on detecting code clones across different programming languages, which is a challenging task since code written in different languages may have varying syntax and structure, even if they serve the same purpose. To address this challenge, researchers have proposed models that can detect code clones across different programming languages \cite{tao2022c4,nafi2019clcdsa,ankali2021detection}. The difference between CLCCD and our research on unseen languages is that CLCCD studies detect clones in various languages, but the models are trained and tested on a \textit{combined} set of programming languages. Therefore, if we intend to use the same model to detect clones in a different programming language that was not included in the training set, we must create a new dataset from scratch. This limits the model's ability to generalize to \textit{unseen} programming languages\footnote{Thus, we do not focus on CLCCD and our focus is only on CCD in this work. It is worth mentioning that applying the same techniques can be interesting to assess the ability of the CLCCD models, but this is out of the scope of our work, due to the high number of required experiments and limited computational resources.}. 
\textcolor{black}{It is worth noting that the studies that address CLCCD are much fewer than the ones that investigate code clone detection in one programming language. Moreover, most of these CLCCD methods focus solely on Java, Python, and C\# and overlook other low-resource languages like Ruby \cite{nafi2019clcdsa,perez2019cross}. }

\textbf{Differences with the existing research:} 
The study of code clone detection models in a few-shot setting that we address in our work here is new. Though previous works assessed the performance of their model for different problem sets \cite{yu2019neural} or tackle the cross-lingual code clone detection \cite{tao2022c4,nafi2019clcdsa,ankali2021detection}, the current research differs from our work from a few perspectives. We evaluate the models' ability in a more rigorous way, and when only a few examples (e.g., $15$) are available for training. Unlike CLCCD works, we consider unseen languages as languages that are not seen during the training phase. We also study a combination of both, when both problem sets and languages are new to the model. Additionally, we apply our experiments for Ruby, in addition to well-studied languages, Java and C++. Finally, we apply few-shot learning algorithms and combine contrastive learning with the few-shot algorithm, which both are among the contributions of our work. 



\section{Background}\label{sec:background}

\subsection{Few-Shot Learning}\label{sec: fewshot}

\begin{figure*}
    \centering
    \includegraphics[width=0.5\textwidth]{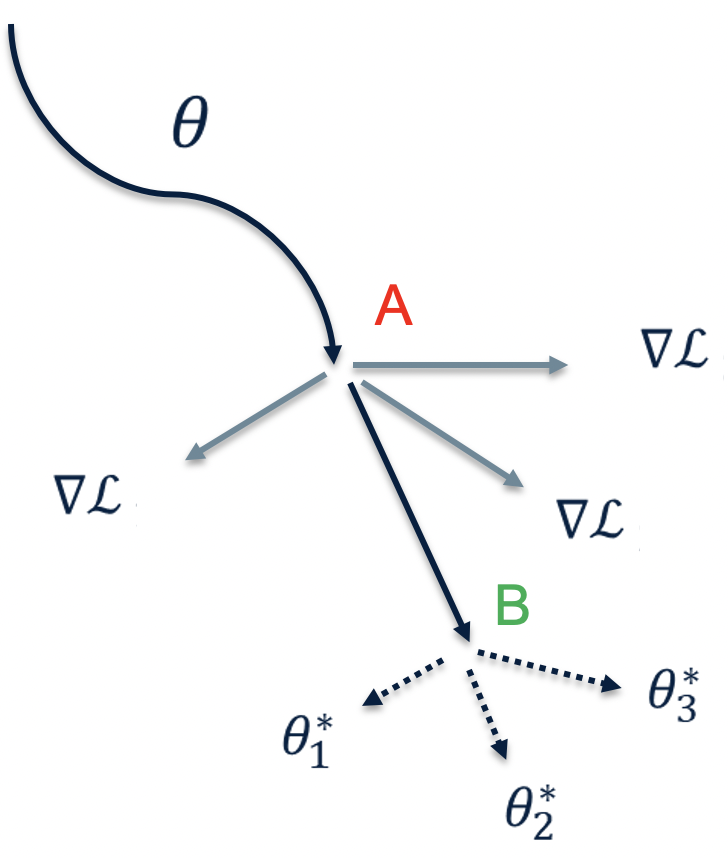}
    \caption{This Figure, which is adopted from \cite{finn2017model}, shows the intuition behind the Model-Agnostic Meta Learning algorithm. Having a model that is parameterized with $\theta$ (A) cannot solve problems in the direction of $\theta^{*}_{1}$,  $\theta^{*}_{2}$, $\theta^{*}_{3}$ accurately. However, if the model is moved to point (B) through meta-learning, it can solve these problems using a few gradient steps.}
    \label{fig:maml}
\end{figure*} 

\begin{figure*}[!h]
    \centering
    \includegraphics[width=0.7\textwidth]{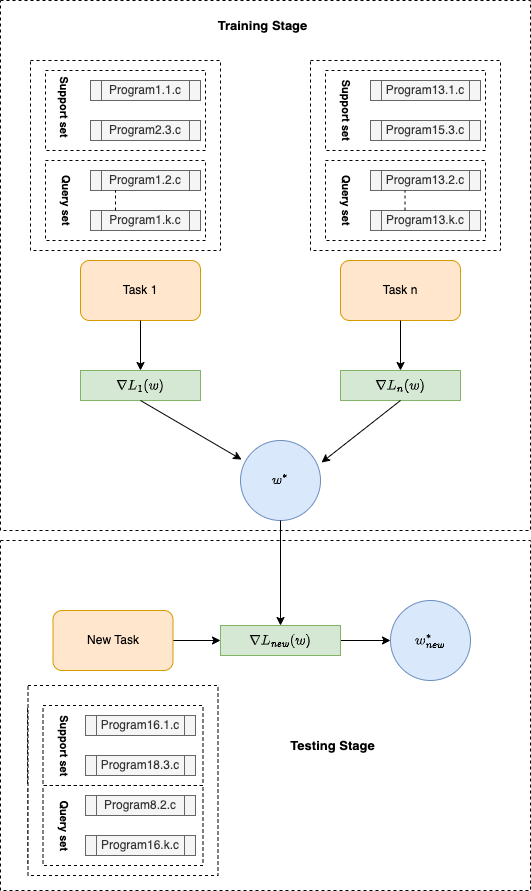}
    \caption{This Figure depicts the process of training in MAML. In MAML, an episode of various tasks is selected randomly in each iteration. Each episode simulates the C-way K-shot learning. This Figure depicts two-way 1-shot episodes. Then, MAML proceeds in two stages. In the inner loop, a base learner is used to compute a loss that is a mid-point for all the tasks in each episode using stochastic gradient steps. Then, a $\nabla L$ is computed using all losses in the inner loop. Finally, an outer loop performs a step of stochastic gradient descent over $\nabla L$. }
    \label{fig:maml-train}
\end{figure*}

Few-shot learning is an area of machine learning where the model is trained using a restricted number of labeled data, sometimes as much as one sample in a downstream task. In this regard, few-shot learning algorithms and models learn a meta learner so that the meta learner can compare the labeled samples, which are called support set, with query examples, for which the model should predict the related class \cite{song2022comprehensive}. 
Figure \ref{fig:maml} shows the intuition behind Model-Agnostic Meta-Learning (MAML), the parametric-based algorithm used in this research for few-shot learning. 
In few-shot learning, we use three different datasets: \textit{train} set, \textit{support} set, and test set (also known as \textit{query} set). 
While the support set and test set share the same label space, the label space of the train set is disjoint with them. 
In this regard, if the support set consists of K samples per each unique class and there exist C classes in the support set, then this problem is called a K-shot C-way \cite{li2021concise}. 
It is possible to learn a classifier using only labeled samples in the support set. However, as the number of labeled data is scarce, the model would be overfitted to a few examples in the support set. Thus, few-shot learning algorithms learn a meta-learner capable of extracting transferable knowledge from the train set so that the model can be successfully fine-tuned using labeled data in the support set and performs more accurately on the test set \cite{li2021concise}.
Accordingly, to better train the meta-learner on the train set, an episode approach similar to \cite{vinyals2016matching} is utilized in most studies to mimic the few-shot learning setting for meta-learning. The process of training MAML is depicted in Figure \ref{fig:maml-train}. An episode is created in each training iteration by sampling C classes in the train set. Then, K-samples are chosen randomly for each class to serve as the support set.
Moreover, a query set is formed using the remaining samples. The model is then trained using the support set in each episode, and the losses for each episode (task) are computed using the query set to perform learning in the inner loop of MAML. Then, in the outer loop of MAML, The model parameters are updated using gradient descent on the aggregate loss computed over all tasks in the episode. The objective is to obtain a model that can quickly adapt to new tasks.
We follow this episode-based approach for code clone detection. 
 The details of how the algorithm is used in our study for few-shot learning are provided in Section \ref{subsec:maml}.

\subsection{Contrastive Learning}

\begin{figure*}
    \centering
    \includegraphics[width=0.85\textwidth]{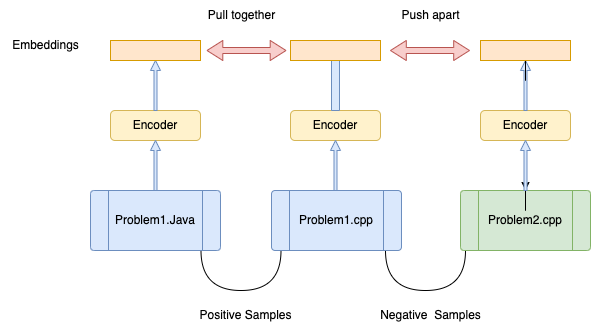}
    \caption{In this Figure, codes from the same problem set are considered positive samples. These positive instances can be from the same or different languages. Code snippets from other problem sets are considered negative examples. The contrastive objective aims to pull the representation of positive samples together and push the embedding of negative ones apart.}
    \label{fig:contrastive}
\end{figure*}

Contrastive learning is a powerful self-supervised approach for unsupervised learning tasks, though it can be used for both supervised and unsupervised learning tasks. This method enables models to learn the representations without needing annotated data. 
In contrastive learning, instead of training a model to predict a ground truth label, the model aims to produce a representation that is similar for positive samples and is different from the negative pairs \cite{le2020contrastive}.
Accordingly, a strategy to generate positive and negative samples to learn the contrastive learning objective should be employed. For instance, ContraCode is a contrastive code representation learning model trained on a dataset of JavaScript code fragments. It utilizes a compiler-to-compiler translator to generate multiple variations of the same code as positive samples. In this way, other code snippets with different variations are considered negative examples in comparison to a given code \cite{jain-etal-2021-contrastive}. 
 In a contrastive supervised learning approach, on the other hand, the objective is defined such that the samples that belong to the same class have similar embeddings. Accordingly, positive samples belong to the same label space, and negative samples have different classes. 
Figure \ref{fig:contrastive} shows how the contrastive objective is trained to build an embedding that is similar for positive examples and different for negative ones.



\section{Methodology}\label{sec:method}

We investigate the answer to our research questions in three scenarios: i) unseen problems, ii) unseen languages, and iii) unseen problems and languages. These scenarios are considered as they shed light on different aspects of the models in terms of their generalizability. 
In CCD, there are two different tasks that we adopt from the literature. The first one is a binary classification CCD used in \cite{yu2019neural,wang2020detecting} and the second one is a retrieval-based CCD used in \cite{ye2020misim,xue2022seed}.
We first conduct all the experiments for binary classification CCD and then based on the results, select experiments for retrieval-based CCD to reduce the number of unnecessary experiments and computation costs.

We conduct our experiments on Java, C++, and Ruby.
Java and C++ are considered as they are widely studied \cite{yu2019neural,wang2021syncobert,GuoRLFT0ZDSFTDC21} and Ruby is selected as it is a low resource language \cite{chen2022transferability} and it has not been studied previously for CCD. As Lei et al. report, $84$ percent of CCD studies focus on Java and C, and the other $16$ percent mainly work on Python, C\#, C++, and Go \cite{LEI2022111141}.
Moreover, Ruby is different from C++ and Java in terms of syntactical structure, and the models cannot benefit from the similarity of the given input languages, thus their ability to detect code clones for languages that are structurally different can be assessed. 
Other programming languages are not considered in our study due to the high number of experiments and limitations in computational resources.

\begin{figure*}
    \centering
    \includegraphics[width=0.9\textwidth]{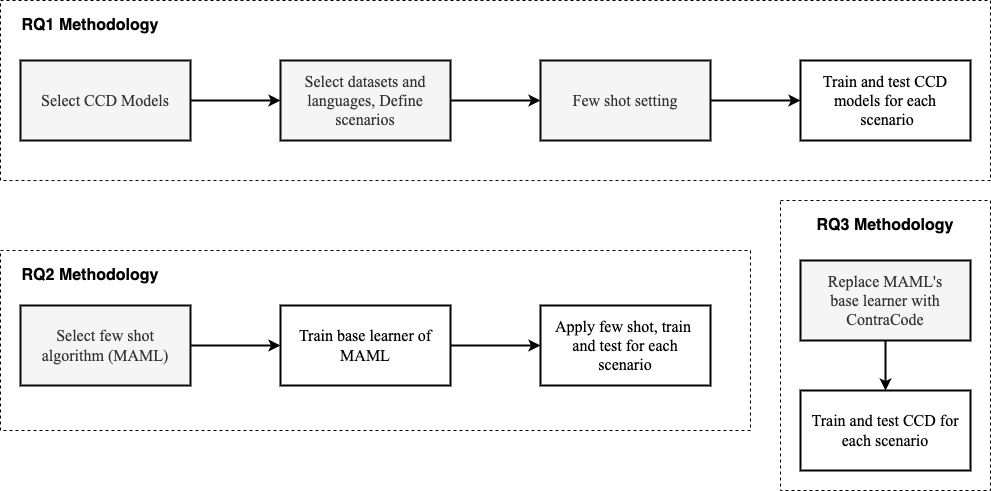}
    \caption{Big picture methodology for each research question.}
    \label{fig:methodology}
\end{figure*}

We selected CodeBERT as a baseline, which has one of the state-of-the-art results for CCD on CodeXGLUE benchmark \cite{lu2021codexglue}. 
Additionally, we used RoBERTa \cite{liu2019roberta} and TBCCD \cite{yu2019neural} to answer RQ1 and RQ2 and ContraCode \cite{jain-etal-2021-contrastive} for RQ3. 
Figure \ref{fig:methodology} demonstrates an overview of our methodology for this research.
First, we evaluate the baseline models in a few-shot setting (RQ1). Then, we apply a widely used few-shot learning approach to evaluate the models in the same settings (RQ2). Finally, we use contrastive learning in the few-shot approach to evaluate the performance of the model in the few-shot setting (RQ3). 

In the rest of this section, we provide the details to answer each research question, the two tasks of code clone detection (i.e., binary classification and retrieval-based), dataset, baselines, and evaluation metrics.

\subsection{Research Questions}\label{sec:RQ}

\textcolor{black}{We answer three research questions in our work, as follows. }
\begin{itemize}
    \item \textcolor{black}{RQ1: What is the performance of the current models for code clone detection in few-shot setting?}
    \item \textcolor{black}{RQ2: What is the performance of few-shot learning method, MAML, for code clone detection?}
    \item \textcolor{black}{RQ3: What is the performance of the few-shot learning models combined with contrastive learning?}
\end{itemize}

In this sub-section, we overview the research questions and our approach to answering them. 
All the RQs are explored for each of the three scenarios. 
To increase the readability of the paper, we provide the overall approach for each RQ here, and then we detail the approach and the answer to each research question for each of the three scenarios in a separate section (Sections \ref{sec:scenario-i}--\ref{sec:scenario-iii}).

\subsubsection{RQ1: Performance of Current Models for Code Clone Detection in Few-Shot Setting}\label{methodRQ1}

In this research question, we investigate the performance of the current models for code clone detection. Following \cite{yu2019neural}, we consider unseen classes as unseen programming problems. For instance, if the model is trained on sorting algorithms, then solutions for string-matching problems are considered an unseen class/problem. For the unseen languages, we consider Java, C++, and Ruby.  

\textbf{Approach:}
To answer RQ1, we fine-tune the baseline models on the training set by using either the first 15 problems of Java, C++, and Ruby for unseen problems or the source language in the case of unseen languages \cite{yu2019neural}. Subsequently, we further fine-tune the models with a few examples from the downstream task in CCD. Finally, we evaluate the models' performance on the test split of the CCD dataset.

For the few-shot setting, we follow the work of {Bansal et al. \cite{bansal2022few}}, and experiment with fine-tuning the models once with 5 samples, once with 10 samples, and once with 15 samples. 
To reduce the number of experiments, we apply this setting for Java and evaluate the best performing models based on the number of samples used for training. This number is then chosen and used for all other experiments. In our work, the models performed the best using 15 samples. So we will use the 15-shot setting for the rest of this study.

\subsubsection{RQ2: Performance of Few-Shot Learning Method for Code Clone Detection} \label{subsec:maml}

After investigating the performance of the current state-of-the-art models for code clone detection in RQ1, we investigate the applicability of few-shot learning models in improving the performance of the models.
In this research question, we study the performance of a popular few-shot learning algorithm, Model Agnostic Meta-Learning {\cite{finn2017model}} for code clone detection. 

\textbf{Approach:} 
The Model Agnostic Meta-Learning algorithm (MAML) {\cite{finn2017model}} is considered to be used for this purpose. 
MAML is chosen as this algorithm is one of the state-of-the-art models for few-shot learning. Moreover, MAML has been applied successfully to natural language processing tasks in previous studies \cite{bansal2022few}. 
In MAML, the training dataset is divided into different tasks. {Each task is an episode, which is formed following the approach introduced in section \ref{sec: fewshot}.} 
Then, the best parameters for each distinct task are calculated using the following equation:

\begin{equation}\label{eq:maml1}
    \theta^{'}_i = \theta - \alpha \nabla_{\theta}\mathcal{L}_{\mathcal{T}_i}(f_{\theta})
\end{equation}

where $f_{\theta}$ is a model parameterized by $\theta$, and $\alpha$ is the learning rate. In this equation, the parameters of the base learner change from $\theta$ to $\theta^{'}$ when the model adapts to task $\mathcal{T}_i$. We use the cross-entropy loss to calculate the loss of each task, following the approach introduced in \cite{bansal2022few}, which is applicable to all experiments except RQ3. 
For RQ3, we use InfoNCE as the loss function since the base learner of MAML will be ContraCode. Then, a query set is formed {based on the episodic approach explained in section \ref{sec: fewshot}} and is used to compute the losses for each task, which are then added together to optimize the meta objective (See Figure \ref{fig:maml-train} for the episodic-based training in the MAML discussed in section \ref{sec: fewshot}). The meta-optimization is calculated using the following equation:

\begin{equation}\label{eq:maml2}
    \theta \longleftarrow \theta - \beta \nabla_{\theta} \sum_{\mathcal{T}_i \sim p(\mathcal{T})} \mathcal{L}_{\mathcal{T}_i} (f_{\theta^{'}_i})
\end{equation}

where $\beta$ is the meta step size. 
In the test time, the model trained using MAML, is first fine-tuned on code clone detection. Finally, the fine-tuned model predicts the label for each example in the test set. 
MAML requires a base learner.
In the inner loop of MAML, any gradient descent-based approach can be employed to compute the losses of each episode. Here, we use this gradient descent-based approach as the base learner of the MAML. TBCCD \cite{devlin2018bert}, CodeBERT \cite{feng-etal-2020-codebert}, and RoBERTa \cite{liu2019roberta} are used as the base learners\footnote{We also used BERT as the base learner of MAML in our experiments. However, the results of the BERT model were very low and therefore, we did not use BERT as the base learner.} of MAML separately\footnote{In an initial assessment of the feasibility of using the models for this study, we tried to use TBCCD as the base learner of MAML as well. However, there is an error while training TBCCD for information retrieval-based CCD tasks, and therefore, we omit this model for RQ2 for retrieval-based CCD tasks.}.     
For all the experiments using MAML, the learning rate for the outer loop is set to $5e-5$ following \cite{bansal2022few}. MAML can learn the inner learning rate as one of its optimization parameters.
According to \cite{bansal2022few}, the internal learning rate is equal to the outer procedure. The outer epochs are set to $100$, and $200,000$ update steps are applied for each episode in training.
In each episode, the support set is created by randomly choosing $15$ samples following \cite{snell2017prototypical}. The remaining instances are used to form the query set to calculate the loss in each training step for each episode.

Using MAML and the base learners as mentioned, we study the three scenarios for this RQ.

\subsubsection{RQ3: Performance of The Few-Shot Learning Models Combined with Contrastive Learning}

\textcolor{teal}{Contrastive learning is used} in \cite{jain-etal-2021-contrastive} and the model learns the functionality of the code fragments, which is important in detecting code clones.
So, here, we investigate whether using a contrastive learning code representation approach can increase the performance of the models in the few-shot setting when integrated with the few-shot learning algorithm. 


\textbf{Approach:}
In this research question, we replace the base learner of MAML with ContraCode {\cite{jain-etal-2021-contrastive}}, a contrastive learning model for code representation learning. The InfoNCE objective function of ContraCode is used in both the inner and outer loops of the Model-Agnostic Meta-Learning algorithm. Unlike previous experiments for RQ2, in the inner loop, a representation is learned for each episode instead of computing a loss with cross-entropy loss. The outer loop optimizes the representation over all episodes. We evaluate the performance of this approach in all three scenarios, unseen problems, unseen languages, and a combination of both, using the same setting as RQ2. 
As a result, by replacing the base-learner with ContraCode,  we will replace the meta-objective function of MAML with InfoNCE {\cite{aitchison2021infonce}}, which is a contrastive learning objective function. 
InfoNCE converts the representation learning into a binary classification so the model predicts the positive pairs between a batch of negative samples. The following equation shows the loss function of InfoNCE:

\begin{equation}
\mathcal{L}_{q, k^{+}, k^{-}} = -log \frac{exp(q.k^{+}/t)}{exp(q.k^{+}/t) + \sum_{k^{-}} exp(q.k^{-}/t)}
\label{eq1}
\end{equation}

In this equation, the loss function value is low when the query $q$ is similar to the positive key $k^{+}$. The parameter $t$ is the temperature hyperparameter.

\subsection{Code Clone Detection Tasks}\label{sec:ccd}
Code clone detection is considered differently in various studies, either as a binary classification or a retrieval-based CCD. Therefore, we study both approaches in our work. We conduct experiments for binary classification CCD for all three scenarios. But, based on the obtained results, and to reduce unnecessary calculations, we decided to pursue the retrieval-based CCD only for the third scenario, in which we assess the ability of the models for unseen problems and languages. 

\textbf{Binary Classification CCD:} In this task, the label of code fragments $C_i$ and $C_j$ is $1$ if they are regarded as code clones and $0$ otherwise.
Accordingly, $n$ code samples in the training set can be denoted by $T = \{ (C_i, C_j, y_{i,j}) | i, j \in n, i \neq j   \}$, in which $y_{i,j}$ is $1$ for clone pairs and $0$ otherwise. A code clone detection algorithm based on deep learning aims to find a function $\Phi$ to compare the representation of two code fragments and assign a label to each pair based on their similarity. In most deep learning models, the cosine similarity $s_{i,j}$ is used to compare the embeddings of code samples. 
The equation to calculate this metric for two code fragments is as follows:

\begin{equation}
    s_{i, j} = \frac{\Phi(C_i) . \Phi(C_j)}{|| \Phi(C_i)|| ||\Phi(C_j)||}
\end{equation}

where $s_{i, j} \in [-1, 1]$, $\Phi(C_i)$ is the embedding of $C_i$, and $\Phi(C_j)$ is the embedding of $C_j$. The algorithm compares this similarity with a threshold $\delta$ to map this metric into a label. 

\textbf{Retrieval-Based CCD:} 
This task compares a query code snippet with $k$ code samples and returns top $R$ code fragments with the most semantic similarity \cite{lu2021codexglue,wang2021syncobert}.
Thus, the training set is represented as follows:

\begin{equation}
    (C_i, \{ C_j, ..., C_m \}, \{C_l, ..., C_p\} | j-m =k \; and \; p-l = R) 
\end{equation}

where $C_i$ is the query code, $C_j, ..., C_m$ are candidates compared to the query code, and $C_l, ..., C_p$ are $R$ detected clones. The parameter $k$ is the number of candidates, and $R$ is the number of the top most similar code fragments to $C_i$ in terms of semantic similarity.
The cosine similarity is the metric used to compare code fragments' representation in this task. 

It is worth noting that retrieval-based code clone detection is very similar to code clone search. However, code clone search has different objectives and requirements. Code clone search is a branch of code clone detection that focuses on developing search engines that aim to look for clones of a query code fragment in a large corpus of code snippets \cite{Keivanloo2021}. Studies on code clone search focus on scalability, response time, and ranking the result set \cite{Keivanloo2021}. However, retrieval-based code clone detection intends to compare the query code with a set of code candidates pairwise and returns a fixed number of most similar matches as code clones. 
In our study, we do not investigate the code clone search and only focus on the code clone detection, either as binary classification or retrieval-based CCD.

\subsection{Dataset}\label{data}

In this study, we extract data from the CodeNET dataset \cite{NEURIPS_DATASETS_AND_BENCHMARKS2021_a5bfc9e0} for all three languages. 
The CodeNet project is released by IBM and composes a high-quality dataset of $14$ million code examples in $50$ programming languages, each aiming to solve one of the $4,000$ problems in Project CodeNet \cite{NEURIPS_DATASETS_AND_BENCHMARKS2021_a5bfc9e0}. 
These problems are gathered from different online resources like AIZU Online Judge and Atcoder websites and contain solutions to the same problems in multiple languages. Thus, this makes it suitable for our experiments, as we can have \textit{multiple} solutions to the \textit{same} problem in \textit{different} programming languages. Henceforth, for each of the scenarios, we can reduce the bias by studying the effect of only one change. 
In other words, for the same set of problems, we can study the performance of models for unseen languages (scenario II); and for different languages, we can study the performance of the models for new problems (scenario I). 

The implementations for each problem are gathered in a specific directory.
The solutions in the same directory can be interpreted as the codes from the same clone, as they provide solutions to the same problem. We use CodeNet to build the dataset for Java, C++, and Ruby and use it to answer RQ1--RQ3 for all three scenarios: unseen problems, unseen languages, and both.
To maintain consistency across different programming languages, we \textcolor{black}{generate datasets with the same} \textit{schema} of two widely used CCD benchmarks, BigCloneBench \cite{svajlenko2016bigcloneeval} for binary code clone detection, and POJ-104 {\cite{yu2019neural,feng-etal-2020-codebert,wang2020detecting,wei2017supervised}} for retrieval-based code clone detection. \textcolor{black}{In this regard, we curate datasets that are similar to BigCloneBench and Poj-104 in terms of file format, i.e., using JSONL, and data fields in those files.}
We extract the data for each of Java, C++, and Ruby from CodeNET as explained below.

\textbf{Data Extraction:} 
The raw data for CodeNet consists of directories and files. Each directory collects all the implementations for a specific problem, and each problem is indicated by an index. The name of the folder is the same as the index of the problem. 
\textcolor{black}{Additionally, CodeNET includes CSV meta-data files that provide information for all submissions related to each problem. Each submission has a unique ID, and the metadata includes various details about them, such as the problem they address, the programming language used, the submission status (which can be either 'accepted' or 'wrong answer'), as well as specifics about time complexity, code size, and memory usage.} 
\textcolor{black}{We first utilized the meta-data to filter out submissions with wrong answers from the problems in CodeNet. Subsequently, we processed the CodeNet data for binary and retrieval-based CCD, as detailed below.}

\begin{table}[]
    \caption{Statistics of binary classification CCD retrieval-based CCD}
    \centering
    \begin{tabular}{@{}lccc@{}}\toprule
         & \# of Samples (Binary)  & \# of Samples (Retrieval)\\
         \midrule
        \textbf{Train} & 1,000,000  &  32,000\\
        \textbf{Dev} & 500,000 &  8,000\\
        \textbf{Test} & 500,000 &  12,000\\
        \bottomrule
    \end{tabular}
    
    \label{tab:datastats}
\end{table}

There are over $1,492$ problems for which the solutions exist in Java, C++, and Ruby. The collected data is in JSON format. The splits for train, dev, and test follow the configurations in BigCloneBench for binary code clone detection and POJ-104 for the code retrieval task. Table \ref{tab:datastats} shows these configurations in our study.


\textbf{Binary Classification Data:} 
For generating a JSON file useful for binary clone detection, we follow the proposed schema for BigCloneBench {\cite{svajlenko2021bigclonebench}}. Accordingly, the JSON file includes:

    \texttt{func1}: the source code for the first function.

    \texttt{func2:} the source code for the second function.

    \texttt{index:} the index of the problem.

     \texttt{id1:} \textcolor{black}{index} of the first function.

   \texttt{id2:} index of the second function.

    \texttt{label:} the label of the sample, either TRUE or FALSE, indicating whether \texttt{func1} and \texttt{func2} are clones of each other.

 
 For extracting the binary classification data from CodeNet, each code sample in each directory is paired with another code snippet from the same folder for generating positive samples (i.e., they have TRUE labels), as all the codes for the same problem share the same semantics.
 As a result, they can be considered as clones of each other and it will be type IV clones, where the code fragments are semantically the same {\cite{ain2019systematic}}. 
For creating the negative samples (i.e., they have FALSE labels), each source code is paired with other code snippets in other folders; as the implementations that belong to different problems are not clones of each other. 
We limit the number of positive and negative samples such that the number of generated instances is the same as BigCloneBench \cite{svajlenko2014towards} and the collected data has balanced classes, following \cite{svajlenko2014towards}. 

As the few-shot algorithm used in this research follows an episode-based approach to training meta-learner, having sequential labeling in the dataset overfits the meta-learner. \textcolor{black}{Sequential labeling occurs when all positive samples or negative samples are grouped together for training in each episode. Thus, sequential labeling creates episodes that only have positive samples or negative samples. This will overfit the base learner that is trained on the subsequent episode, which affects the MAML training.}
To avoid this issue, we shuffle the dataset randomly. 
For the evaluation of unseen problems, the same number of samples as shown in Table \ref{tab:datastats} are selected from the first $15$ problems for both the training and development sets. 
\textcolor{black}{Note that although the first 15 problems are used to generate samples for training and development, the sample pairs in these sets (i.e., training and development) are different, as they contain pairs of codes, each pair being clones or non-clones.}
Finally, each problem set is tested using a set of $500,000$ instances as the test set. \textcolor{black}{Each of these instances consists of clone or non-clone code snippet pairs.}
We refer to this dataset as CodeNet\textsubscript{B} in the rest of the paper, but for simplicity, we do not mention the specific subset language of CodeNet\textsubscript{B} in each experiment.

\textbf{Code Retrieval Data:} 
For extracting this dataset from CodeNet, the same raw data used for binary clone detection is considered. The JSON files generated for train, dev, and test follow the schema of POJ-104 in the Hugging Face\footnote{{https://huggingface.co/}} repository. Each JSON file includes the following fields:

\texttt{code:} the source code of the current sample.

\texttt{id:} the index of the current instance.

\texttt{label:} the index of the problem that the current sample aims to solve. 

We wrote a script to traverse all directories for each problem. First, the folder's name is assigned to the label for the current instance. Then, the algorithm processes all the submissions for the current directory and extracts their source code to be set as the code field of the data. The dataset includes $105$ problem sets for each language, each consisting of $500$ submissions. In the experiments of this paper, the training and development sets are selected from the first $15$ problems, with $12$ problems being chosen for the training set and three problems serving as the development sets for training baselines. For calculating the performance of models on this dataset, we use MAP@R, which is described in section \ref{evaluationMetircs}. It is worth noting that the default value of the R in this dataset is $499$.


We refer to this dataset as CodeNet\textsubscript{R} in the rest of the paper, but for simplicity, we do not mention the specific subset language of CodeNet\textsubscript{R} in each experiment.

It is worth mentioning that the problems and solutions in CodeNet\textsubscript{B} and CodeNet\textsubscript{R} are the same, meaning that the records are the same and only the data is organized differently in these datasets to address our needs for binary classification or clone retrieval in our study.

\subsection{Baselines}

\textbf{CodeBERT} is a bimodal pre-trained model for program understanding. CodeBERT is trained to learn a representation that is useful for programming language and natural language processing downstream tasks like code search and code documentation generation. 
CodeBERT employs a transformer based-architecture and it pre-trains a hybrid objective function for detecting replaced tokens. This model could achieve near the state-of-the-art performance in many downstream tasks like code search and code documentation generation \cite{feng-etal-2020-codebert}. Moreover, it has been used as the baseline in previous studies for code clone detection \cite{yu2023graph,zhang2023efficient} and is the top model in the CodeXGLUE leaderboard\footnote{https://microsoft.github.io/CodeXGLUE/}. Thus, we chose it as a baseline.

\textbf{RoBERTa} is an improved version of the BERT model, and uses a Masked Language Model objective for training. This allows RoBERTa to learn the intentionally hidden tokens in the text, providing a deeper understanding of language patterns and relationships. Unlike BERT, RoBERTa ignores the Next Sentence Prediction objective and instead focuses solely on the Masked Language Model objective.
Additionally, RoBERTa uses a larger learning rate and mini-batches during training, allowing it to converge faster and obtain better results. RoBERTa also uses a dynamic masking strategy, which allows it to make use of more of the training data and improve its generalization ability. Finally, RoBERTa is trained on a much larger corpus of text, providing it with a broader knowledge of language patterns and relationships \cite{liu2019roberta}. Since RoBERTa has been employed as a baseline in previous studies for code clone detection \cite{wang2021syncobert,lu2021codexglue}, and is in the CodeXGLUE leaderboard\footnote{https://microsoft.github.io/CodeXGLUE/}, we use this model as a baseline in our research.

\textbf{TBCCD} is designed for code clone detection \cite{yu2019neural}. TBCCD proposes tree-convolution to detect semantical clones by combining the lexical and structural information. The structural information is extracted from the Abstract Syntax Tree (AST) of source codes, and the lexical information is captured from source code tokens. TBCCD employs a siamese architecture that consists of two deep neural networks. In each of these neural networks, a tree-based convolution is applied to the AST of query source codes. Then, a max pooling layer is applied to the embeddings produced in the previous step, and the result is used in a fully connected layer. In this research, we use the standard version of TBCCD, excluding the position-aware character embedding (PACE), for a fair comparison with other baseline architectures. 

\textbf{ContraCode} pre-trains a neural network using a contrastive objective \cite{jain-etal-2021-contrastive}. Thus, this model learns to identify functionally similar code snippets. ContraCode employs a source-to-source compiler to fabricate the positive variations of each code fragment. As a result, each variation of other code snippets is considered a negative counterpart. Then, ContraCode uses a contrastive objective to create a representation that is similar for positive examples and dissimilar for negative ones. ContraCode has been successfully applied to different code understanding tasks, including Code Clone detection. \\

\subsection{Evaluation Metrics}\label{evaluationMetircs}

\subsubsection{F1-Score}

F1-Score, also known as the F-Score or F-Measure, is a commonly used performance metric for evaluating the accuracy of a binary classifier. It is a balance between precision and recall. Precision measures the proportion of true positive predictions out of all positive predictions made by the classifier, while recall measures the proportion of true positive predictions out of all actual positive instances. The F1-Score is computed as:

\begin{equation}\label{eq:f1-score}
F_1 = 2 \cdot \frac{precision \cdot recall}{precision + recall}
\end{equation}

where precision and recall are defined as:

\begin{equation}
precision = \frac{True Positives}{True Positives + False Positives}
\end{equation}

\begin{equation}
recall = \frac{True Positives}{True Positives + False Negatives}
\end{equation}

with a perfect score of $1.0$ indicating perfect precision and recall, and a score of $0.0$ indicating the worst possible performance. In this paper, we use F1-Score for computing the performance of models for binary code clone detection.

\subsubsection{MAP@R}

Mean Average Precision at rank R (MAP@R) is a commonly used evaluation metric in information retrieval. It measures the average precision of a system's top R-ranked items for a set of queries or items. Equation \ref{eq:map} computes the MAP, where $Q$ is the total number of queries in the evaluation dataset, and $AP_i$ represents the average precision for the $i-th$ query. 

\begin{equation}\label{eq:map}
MAP = \frac{1}{Q} \sum_{i=1}^{Q} AP_i
\end{equation}

As a result, MAP@R computes the precision for each query code snippet. Precision is defined as the proportion of relevant code candidates in the top R-ranked codes. It is calculated for each query code fragment, and then the average is taken over all queries. MAP@R provides a summary of the precision scores for all queries and gives an overall measure of the system's performance.

\section{Scenario I: Unseen Problems}\label{sec:scenario-i}

In this section, we will discuss the approach, experimental setup, and the results for unseen problems for binary classification code clone detection. 

\subsection{Approach}

\begin{figure*}
    \centering
    \includegraphics[width=0.9\textwidth]{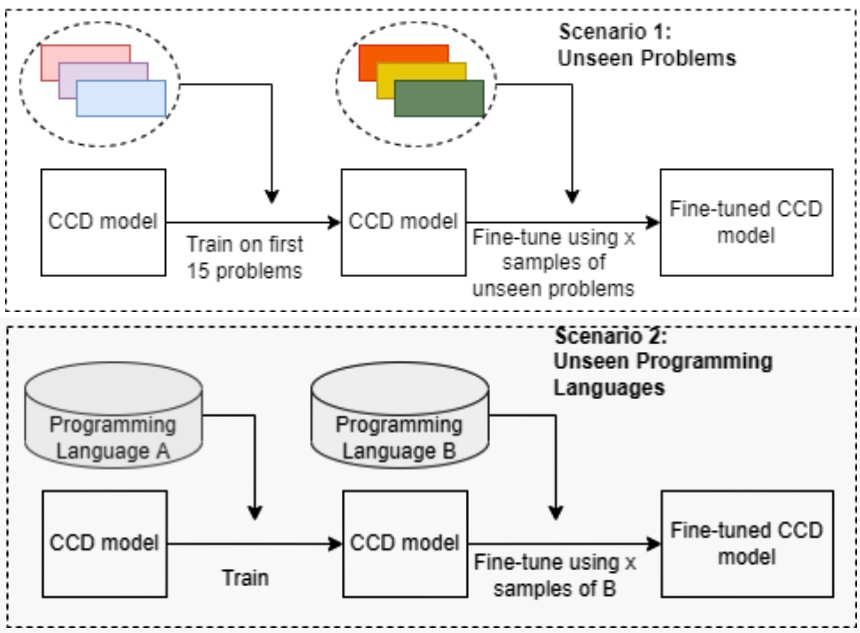}
    \caption{\textbf{Top:} Training process for unseen problems. The dataset is split into groups of 15 problem sets. The first 15 problems are used for training the model, and then the model is fined-tuned using a few (5,10, or 15) samples of unseen problems and tested. \textbf{Bottom:} Training process for unseen programming languages. First, the model is trained on programming language A and then a few (5, 10, 15) records of a new programming language (B) are shown to the model, and the model is tested on the new programming language.}
    \label{fig:training}
\end{figure*}

The steps for training and testing the models in case of unseen problems are shown in Figure \ref{fig:training}. In this scenario, the language is kept the same and the problems vary.
We use the CodeNet\textsubscript{B} and CodeNet\textsubscript{R} datasets. 

We followed the work of Yu et al. \cite{yu2019neural} for the settings of this experiment.
We only applied the experiments on $105$ problems randomly selected from the $1,492$ problems, for each language separately. The problems are divided into groups of $15$ classes randomly, leading to $7$ groups in total. 
We fully trained/fine-tuned each of the baseline models separately on the first $15$ problems (the first group with IDs $1-15$). For the few-shot setting, we then fine-tune the trained model using a few samples from the second group of problems (IDs $16-30$) and then test on the rest of the problems in this group.
This process is repeated for each of the problem groups with IDs $31-45$, $46-60$, $61-75$, $76-90$, and $91-105$ separately, where the model is trained on the first $15$ problems, fine-tuned using a few samples of each of these groups and then tested on that group.
As explained in Section \ref{methodRQ1}, the `few' here refers to $5$, $10$, and $15$ samples in the case of Java and $15$ samples for all other experiments. 
For the $15$-shot part, we chose \textit{one} code snippet from \textit{each} problem in the set. This decision was motivated by previous research, which has shown that the main reason for the poor performance of models on new problems is the presence of unseen tokens \cite{yu2019neural}. By selecting one code fragment from each problem in the set, we aim to increase the likelihood of the model encountering different tokens in the set.

For unseen problems, following \cite{yu2019neural} that consists of $499$ code fragments for each problem, we include the same number of code fragments per problem in each language. 
To address RQ2, we employed MAML and used CodeBERT, RoBERTa, and TBCCD as the base learners of the model. For RQ3, we used ContraCode as the base learner of MAML. ContraCode was pre-trained on Javascript code snippets. To generate positive and negative samples, they use a compiler-to-compiler method.
While the ContraCode is pre-trained using Javascript code fragments, the transformer version can be fine-tuned on other programming languages. So, in our experiments, we first fine-tuned the model using the first $15$ problems of the downstream programming language. In the meta-training phase, ContraCode learns the embedding of different episodes, which consists of code fragments of the downstream language.

\textbf{Experimental Setup}
For CodeBERT and RoBERTa, we used a learning rate of $5e-5$ and l2-norm with a maximum gradient norm of $1.0$ to avoid overfitting, as recommended by the authors of the model. Each of these models are trained for two epochs following the configurations provided in the CodeXCLUE. 
For TBCCD, we set the learning rate to be $0.0002$, and we trained the model for ten epochs for all experiments. 

To run the few-shot algorithm, MAML, the $\alpha$ and $\beta$ hyperparameters need to be set. 
The $\alpha$ is the learning rate for training the base learner of the algorithm. The second parameter is the meta-learning rate $\beta$, which is used for training the meta-learner. 
Following the work of Bansal et al. \cite{bansal-etal-2020-learning}, we set $\alpha$ to $1e-05$, and $\beta$ to $1e-05$. In creating episodes for training MAML, we used $15$ samples as the support set for the meta-learner. To minimize the number of unseen tokens, each of these $15$ code snippets is randomly selected from unique problems in the training set. Then, the model is trained for ten epochs. Due to restrictions in the computational resources, and a large number of samples (one million code samples), we did not increase the number of epochs.

\begin{table}[]
    \centering
    \caption{The results of CodeBERT, RoBERTa and TBCCD in few-shot learning setup for unseen problems (\textit{first scenario)} for \textbf{Java}, for binary classification of code clones. The models are fully trained on the first 15 problems and then tested in few-shot setup on the other problem sets. The numbers are the F1-Scores for 5-shot, 10-shot, and 15-shot settings, respectively, separated by ``/''. The best scores are bold.}
    \resizebox{0.85\columnwidth}{!}{%
    \begin{tabular}{@{}lccc@{}}\toprule
   Problem \# & CodeBERT & RoBERTa & TBCCD\\
    \midrule

        P \#16-30 & 45.7/\textbf{48.1}/48.5  & 41.2/\textbf{46.4}/47.1 & 47.5/\textbf{49.5}/49.4\\
        
        P \#31-45 & 48.7/50.4/\textbf{51.2} & 43.1/47.9/\textbf{49.9} & 49.7/51.3/\textbf{52.1}\\
        
        P \#46-60 &44.6/46.9/\textbf{50.9} & 40.9/43.6/\textbf{47.3} & 46.4/49.1/\textbf{52.9}
        \\
        
        P \#61-75 &48.1/\textbf{49.4/}48.3 & 43.9/\textbf{47.8}/47.8 & 49.3/\textbf{51.3}/49.3
        \\
        
        P \#76-90 & 45.9/47.6/\textbf{51.6} & 45.2/47.1/\textbf{50.2} & 50.4/49.2/\textbf{52.8}
        \\
        
        P \#91-105
        & 49.8/\textbf{51.2}/49.6 & 47.3/\textbf{50.6}/47.7 & 50.1/\textbf{51.8}/49.2
        \\

    \bottomrule
     Mean & 47.13/48.93/\textbf{50.01} & 43.6/47.2/\textbf{48.3} & 48.9/50.36/\textbf{50.95}
    \end{tabular}}
    
    \label{tab:rq1-sc1-java}
\end{table}

\subsection{Results}

In this section, we report the results for the binary classification task in all the research questions for the unseen problems. 

\subsubsection{RQ1: Performance of Models in Few-Shot Setup}

Table \ref{tab:rq1-sc1-java} shows the results of CodeBERT, RoBERTa, and TBCCD trained in a few-shot setting for binary classification CCD in Java. 
The reported numbers are F1-Scores for $5$-, $10$-, and $15$-shot, where the scores are separated by a `/', and the highest scores are shown in bold. As the mean scores show, the average F1-Scores are in the high $40$s for CodeBERT and RoBERTa and reach $50$ for TBCCD. 
This is compared to having scores of $93.7$ for CodeBERT and $85.6$ for RoBERTa when the model is fully fine-tuned on the Java portion of $CodeNet_{B}$, instead of being used in a few-shot setup. TBCCD also achieved $96.2$ F1-Score for clone detection in Java when it is fully fine-tuned on $CodeNet_{B}$ Java subset. \textcolor{black}{When we apply the models in 15-shot setting, the F1 scores drop significantly. For example, the average F1 score of TBCCD in 15-shot setting for all the problem sets (see Table \ref{tab:rq1-sc1-java}) is $50.95$ and the score has reduced by $(96.2-50.95=)45.24$.} 


\begin{table}[]
    \centering
    \caption{The results of the models for a few-shot setup and when MAML is applied for \textbf{C++} in the\textit{ first scenario (unseen problems)} for binary classification CCD. The F1-Scores are reported when the models are tested on sets of unseen problems.
    The first number is the results of CodeBERT, RoBERTa, and TBCCD in a 15-shot learning setup for unseen problems, followed by the results of applying MAML. The numbers in the two experiments are separated by ``/''. The number in the parentheses shows the improvement in F1-Score. The last column shows the results for ContraCode (RQ3). }
    \resizebox{0.95\columnwidth}{!}{%
    \begin{tabular}{@{}lcccc@{}}\toprule
   Problem \# & CodeBERT& RoBERTA & TBCCD & ContraCode\\
    \midrule

        P \#16-30 & 29.6/52.3 (\textbf{22.7}) & 27.1/48.3 (\textbf{21.2}) & 25.6/47.3 (\textbf{21.7}) & 47.5 \\
        
        P \#31-45
        & 31.7/53.7 (\textbf{22}) & 27.3/47.7 (\textbf{20.4}) & 33.2/52.8 (\textbf{19.6}) & 50.4\\ 
        
        P \#46-60
        &27.3/49.8 (\textbf{22.5}) & 25.5/46.1 (\textbf{20.6}) & 29.7/50.6 (\textbf{20.9}) & 47.3
        \\
        
        P \#61-75
        &30.8/53.4 (\textbf{22.6}) & 24.6/45.7 (\textbf{21.1}) & 30.8/51.4 (\textbf{20.6}) & 51.6
        \\
        
        P \#76-90
        & 29.5/49.1 (\textbf{19.6}) & 23.9/46.5 (\textbf{22.6}) & 33.5/52.7 (\textbf{19.2}) & 48.9
        \\
        
        P \#91-105
        & 26.9/47.3 (\textbf{20.4}) & 28.3/48.6 (\textbf{20.3}) & 30.2/52.1 (\textbf{21.9}) & 50.2
        \\

    \bottomrule
    Mean & 28.80/50.93 (\textbf{22.13}) & 24.95/47.15 (\textbf{22.2}) & 30.50/51.4 (\textbf{20.9}) & 49.32
    \end{tabular}}
    
    \label{tab:unseenproblems-bccd-cplusplus}
\end{table}

As the best scores mostly belong to the models when they are trained with $15$ samples, we use $15$-shot training for other languages and all other experiments. 
Tables \ref{tab:unseenproblems-bccd-cplusplus} and \ref{tab:unseen_problems_bccd_ruby} report the F1-Scores of the baselines in $15$-shot setting for C++ and Ruby, respectively, in binary classification CCD.
\textcolor{black}{The numbers in Tables \ref{tab:unseenproblems-bccd-cplusplus} and \ref{tab:unseen_problems_bccd_ruby} that represent the results for this RQ are the first number before ``/'' in the CodeBERT, RoBERTa, and TBCCD columns. The numbers after ``/'' are the results of RQ2 that we explain in the next subsection. }
The numbers are in mid-$20$s to $30$ for both languages, and similar to Java, the results of the TBCCD are higher than the other ones. 
We observe that the scores for Java are relatively higher than the ones for C++ and Ruby in this scenario even when CodeBERT is used. The reason can be that CodeBERT is pre-trained on programming languages with a dominant number of samples from Java. 
C++ is not used in the pre-training of CodeBERT, and only $97,580$ training samples of  Ruby are used in the pre-training data compared to $2,070,643$ of java code samples \cite{feng-etal-2020-codebert}.
Interestingly, RoBERTa's scores are close to CodeBERT. 
We might relate the high scores of RoBERTa for Java to the fact that Java is a high-level language and has some APIs, which are similar to human language.
For all three languages, TBCCD has the highest scores among the models, which we relate to being specifically designed for code clone detection. Therefore, its design could benefit in obtaining higher scores.


While experiments in this research consist of CodeNET samples, one can see the vulnerability of baselines in the face of unseen problems in a few-shot setting. 
Note that TBCCD can achieve an F1-Score of $99$ percent on a binary classification version of POJ-104, which consists of C and C++. Again, comparing the results of this research with previous results achieved using TBCCD indicates the weakness of baselines when evaluate in a few-shot learning set-up.

\begin{table}[]
    \centering
    \caption{The results of the models for a few-shot setup and when MAML is applied for \textbf{Ruby} in the\textit{ first scenario (unseen problems)} for binary classification CCD. The F1-Scores are reported when the models are tested on sets of unseen problems.
    The first number is the results of CodeBERT, RoBERTa, and TBCCD in a 15-shot learning setup for unseen problems, followed by the results of applying MAML. The numbers in the two experiments are separated by ``/''. The number in the parentheses shows the improvement in F1-Scores. The last column represents the results of ContraCode (RQ3). }
    \resizebox{0.95\columnwidth}{!}{%
    \begin{tabular}{@{}lcccc@{}}\toprule
   Problem \# & CodeBERT& RoBERTA & TBCCD & ContraCode\\
    \midrule

        P \#16-30 & 24.2/46.3 (\textbf{22.1}) & 21.3/47.7 (\textbf{26.4}) & 23.6/46.1 (\textbf{22.5}) & 40.6 \\
        
        P \#31-45
        & 21.9/43.5 (\textbf{21.6}) & 22.6/42.1 (\textbf{19.5}) & 24.2/45.6 (\textbf{21.4}) & 41.4\\ 
        
        P \#46-60
        &22.5/47.6 (\textbf{25.1}) & 19.9/44.3 (\textbf{24.4}) & 26.1/43.2 (\textbf{17.1}) & 39.3
        \\
        
        P \#61-75
        &19.2/43.6 (\textbf{24.4}) & 20.3/41.7 (\textbf{21.4}) & 22.8/46.1 (\textbf{23.3}) & 38.9
        \\
        
        P \#76-90
        & 21.2/47.2 (\textbf{26.0}) & 20.4/45.5 (\textbf{25.1}) & 24.1/46.8 (\textbf{22.7}) & 40.5
        \\
        
        P \#91-105
        & 19.7/43.7 (\textbf{24.0}) & 18.7/41.3 (\textbf{22.6}) & 24.3/43.6 (\textbf{19.3}) & 39.2
        \\

    \bottomrule
    Mean & 21.43/45.31 (\textbf{23.88}) & 20.53/43.26 (\textbf{23.73}) & 24.18/45.23 (\textbf{21.05}) & 39.98
    \end{tabular}}
    
    \label{tab:unseen_problems_bccd_ruby}
\end{table}

\subsubsection{RQ2: Performance of Models with Few-Shot Algorithm}

To boost the performance of baselines in a few-shot learning setup, we evaluated the MAML algorithm. 
We reported the results of binary classification CCD when MAML is applied in Tables \ref{tab:unseenproblems-rq2-3-java}, \ref{tab:unseenproblems-bccd-cplusplus}, and \ref{tab:unseen_problems_bccd_ruby} for Java, C++, and Ruby, respectively. \textcolor{black}{ In these tables, the numbers before ``/'' are the ones before using MAML. The numbers after ``/'' are the results of RQ2, when MAML is used. We also added the improvements to the F1-scores in parenthesis in these tables for better comparison. The last column is used for RQ3, so we only discuss the results in columns CodeBERT, RoBERTa, and TBCCD here.}
We trained the meta-learner in MAML using the first 15 problems. Then, 15 samples from each unique problem in each problem set are chosen randomly and are employed as the support set. The model is then evaluated for each problem set separately.

\begin{figure*}
    \centering
    \includegraphics[width=1.1\textwidth]{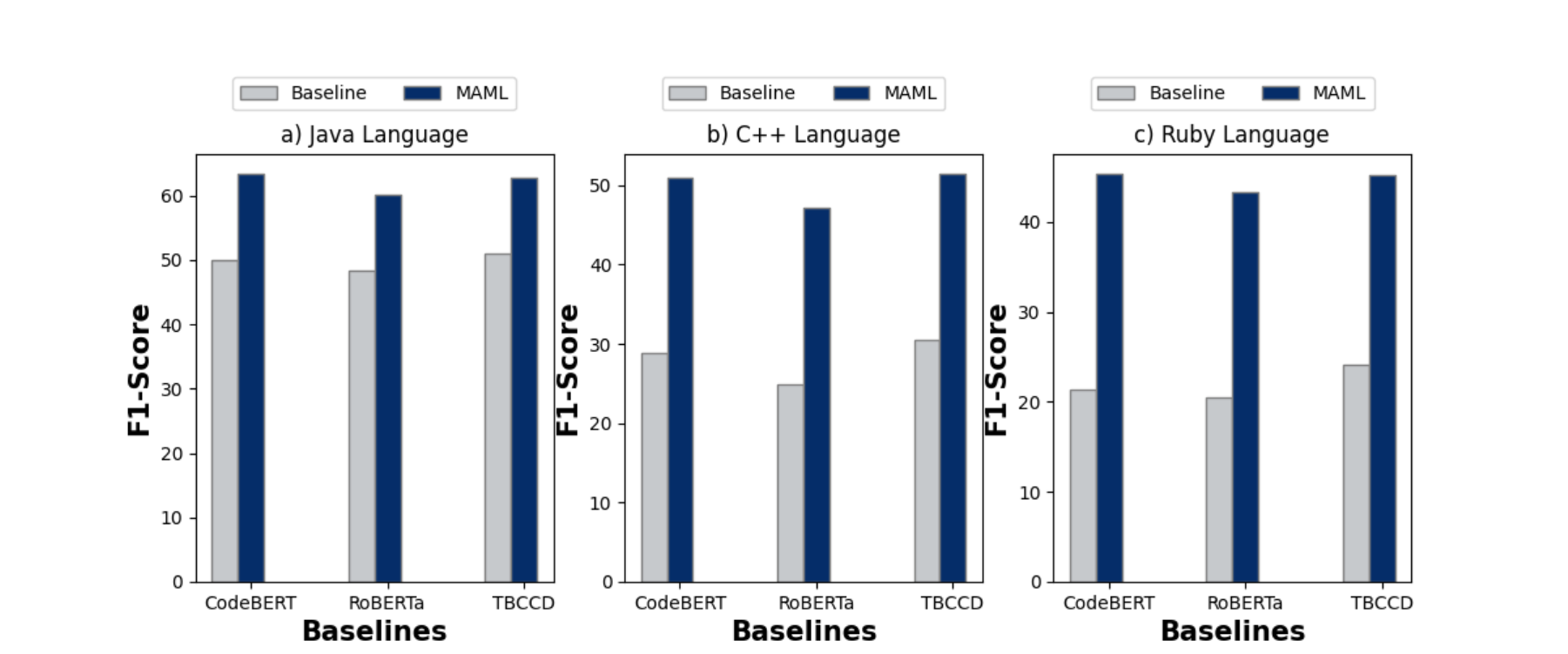}
    \caption{Comparison of the F1-scores for CodeBERT, RoBERTa, and TBCCD, with and without using MAML in a few-shot setting for the first scenario of binary code clone detection. Each plot shows the results of the models for one of the languages, the left one is on Java, the middle one for C++, and the right one is on Ruby.
    } 
    \label{fig:compare_unseen_binray}
\end{figure*}

\begin{figure*}
    \centering
    \includegraphics[width=1.1\textwidth]{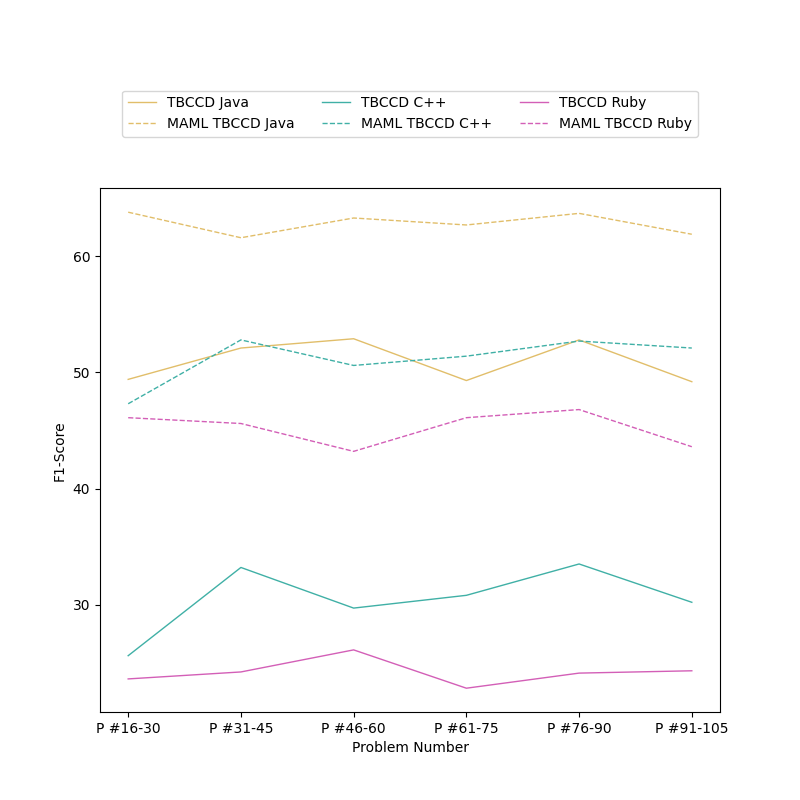}
    \caption{F1-score of TBCCD in the first scenario for binary CCD without (solid lines) and with MAML (dashed lines) for each of the problem sets. The colors indicate the plots for different programming languages. In all languages and all problem sets, higher F1-scores are achieved when MAML is used. }
    \label{fig:TBCCDunseenbinary}
\end{figure*}

Figure \ref{fig:compare_unseen_binray} compares the F1-Scores of all models in the few-shot setting without MAML (\textcolor{gray}{gray} plots) and when MAML is applied (\textcolor{Blue}{blue} plots). 
For Java, MAML improves the F1-Score from $50$ to $\sim60$ for all models. The improvement in C++ and Ruby with MAML is more significant, where MAML boosts the F1 by $20$--$26$ scores. 
For all three languages, the performance of the models is increased significantly by applying the few-shot algorithm. 
On average, MAML can improve the performance of CodeBERT by 13.15 for Java, $22.13$ for C++, and $23.88$ for Ruby.
The enhancement for TBCCD using MAML are $11.88$, $20.9$, and $21.05$, for Java, C++, and Ruby, respectively. Finally, MAML improves the performance of RoBERTa by $11.83$ for Java, $22.2$ for C++, and $23.73$ for Ruby.
Figure \ref{fig:TBCCDunseenbinary} summarizes the results of RQ1 and RQ2 in this section for TBCCD. The dashed-lined plots are the TBCCD plus MAML, and the solid lines show the scores of TBCCD in a few-shot setting without MAML. Each of the three languages are shown with a different color. In all cases, there is a significant improvement in the results when MAML is used.

\begin{table}[]
    \centering
    \caption{The results of the models for few-shot setup and when MAML is applied for \textbf{Java} in the\textit{ first scenario, unseen problems} for binary code clone detection. The F1-Scores are reported when the models are tested on sets of unseen problems.
    The first number is the results of CodeBERT, RoBERTa, and TBCCD in a 15-shot learning setup for unseen problems, followed by the results of applying MAML with different base learners. The numbers in the two experiments are separated by ``/''. The number in the parentheses shows the improvement in F1-scores. The last column shows the results of ContraCode (RQ3).}
    \resizebox{0.95\columnwidth}{!}{%
    \begin{tabular}{@{}lcccc@{}}\toprule
   Problem \# & CodeBERT& RoBERTA & TBCCD & ContraCode\\
    \midrule

        P \#16-30 & 48.5/62.4 (\textbf{13.9}) & 47.1/61.6 (\textbf{14.5}) & 49.4/63.8 (\textbf{14.4}) & 61.2\\
        
        P \#31-45
        & 51.2/65.1 (\textbf{13.9}) & 49.9/60.3 (\textbf{10.4}) & 52.1/61.6 (\textbf{9.5}) & 62.4\\ 
        
        P \#46-60
        &50.9/62.7 (\textbf{11.8}) & 47.3/59.2 (\textbf{11.9}) & 52.9/63.3 (\textbf{10.4}) & 
        61.8\\
        
        P \#61-75
        &48.3/62.1 (\textbf{13.8}) & 47.8/59.4 (\textbf{11.6}) & 49.3/62.7 (\textbf{13.4}) & 
        60.2\\
        
        P \#76-90
        & 51.6/64.9 (\textbf{13.3}) & 50.2/60.8 (\textbf{10.6}) & 52.8/63.7 (\textbf{10.9}) &
        62.3\\
        
        P \#91-105
        & 49.6/61.8 (\textbf{12.2}) & 47.7/59.7 (\textbf{12}) & 49.2/61.9 (\textbf{12.7}) & 
        63.1\\

    \bottomrule
    Mean & 50.01/63.16 (\textbf{13.15}) & 48.33/60.16 (\textbf{11.83}) & 50.95/62.83 (\textbf{11.88}) & 61.83
    \end{tabular}}
    
    \label{tab:unseenproblems-rq2-3-java}
\end{table}

\subsubsection{RQ3: Performance of Models with Contrastive Learning}

The last column of Tables \ref{tab:unseenproblems-rq2-3-java}, \ref{tab:unseenproblems-bccd-cplusplus}, and \ref{tab:unseen_problems_bccd_ruby} show the results of using ContraCode for Java, C++, and Ruby, respectively. 
Though MAML could improve the results of all models by $\sim11-23$ scores, the F1-scores are still in the range of $60$s for Java, $50$s for C++, and low-mid $40$s for Ruby.  
To further improve the results, we replaced the base learner of MAML using the ContraCode \cite{jain-etal-2021-contrastive}, which is a contrastive learning approach.

Interestingly, using ContraCode has a negative impact on the scores compared to when CodeBERT, RoBERTa, or TBCCD are used as base learners of MAML. For all three languages, the scores are slightly reduced or are on par with the previous ones when MAML is used. Using ContraCode has the most negative impact on Ruby, where the average F1-scores are dropped from $45.23$ to $39.98$.



\begin{framed}

\textcolor{purple}{\textbf{Summary of Scenario I for Unseen Problems in Binary CCD}} \\
The results of the binary code clone detection for unseen problems indicate that the baselines are vulnerable when confronted with unseen problem sets. The poor performance of these models can be attributed to the presence of unseen tokens in each problem set. In contrast, MAML significantly improved the performance of models in all languages, and for all problem sets. However, using ContraCode as the base learner did not noticeably improve the performance of MAML when compared to other baselines.

\end{framed}

\begin{table}[]
    \centering
    \caption{The cosine similarity between the first problem and a random problem from other problem sets in Java, C++, and Ruby.}
    \resizebox{0.47\columnwidth}{!}{%
    \begin{tabular}{@{}lccc@{}}\toprule
   Problem \# & Java & C++ & Ruby\\
    \midrule

        P \#18 & 0.929 & 0.781 & 0.479
        \\
        
        P \#38 & 0.886 & 0.759 & 0.486
        \\
        
        P \#52 & 0.878 & 0.701 & 0.656
        \\
        
        P \#67 & 0.909 & 0.759 & 0.532
        \\
        
        P \#81 & 0.888 & 0.776 & 0.716
        \\
        
        P \#102 & 0.896 & 0.673 & 0.516

        \\

    \bottomrule
    Mean &0.897 & 0.741 & 0.564
    \end{tabular}}
    
    \label{tab:similarity}
\end{table}

\subsection{Discussions} \label{subsec:scenario1-discussions}

Yu et al mention that the primary reason for models' poor performance when faced with new problems is the presence of previously unseen tokens in those problem sets \cite{yu2019neural}. We tried to address this by showing one example from each problem when training the models in a few-shot setting. 


To gain a deeper understanding of the performance disparities between the three programming languages, as well as the differences in F1-score across various problem sets in Tables \ref{tab:unseenproblems-rq2-3-java}, \ref{tab:unseenproblems-bccd-cplusplus}, and \ref{tab:unseen_problems_bccd_ruby}, we calculated the cosine similarity between all submissions of samples from problem one and all submissions from each of the problem sets in the test set. As there are a large number of problems, we randomly chose one problem from each of the six problem sets under study (i.e., problem sets shown in Tables \ref{tab:unseenproblems-rq2-3-java}, \ref{tab:unseenproblems-bccd-cplusplus}, and \ref{tab:unseen_problems_bccd_ruby}). 
The randomly chosen problems are shown in Table \ref{tab:similarity}. For example, problem \#$18$ belongs to the first set of problems with IDs $16-30$ (see Table \ref{tab:unseen_problems_bccd_ruby} as an example), problem \#$38$ is from the second set of problems with IDs $31-45$ and so on. 
We then computed the similarity of each pair of the codes from problem one and the code from the chosen problem and report the average of their cosine similarity, as reported in Table \ref{tab:similarity}.

To calculate the cosine similarities, we used the vector representations of the problems.
We employed the Sentence-Transformer approach with the BERT-Large model \cite{reimers-gurevych-2019-sentence} as the encoder to generate embeddings of code fragments. Sentence-Transformer uses mean pooling to compute the embeddings of each code fragment, ensuring that all code fragments have the same length, which is necessary for computing cosine similarity. Unlike pure BERT, which generates a 768-dimensional representation for each token in a code fragment, Sentence-Transformer resolves the issue of varying length code snippets through mean pooling. To maintain fairness in comparison across different languages, we only considered the code tokens in creating the embeddings, ignoring the code structure.

After extracting the embeddings, we calculated the cosine similarity between each pair of the code fragments with the code snippets from problem one using the following formula: 
$cos(\theta) = \frac{\mathbf{A} \cdot \mathbf{B}}{\left|\mathbf{A}\right| \left|\mathbf{B}\right|} $.

The result is a score between $0$ and $1$, where $1$ indicates the most similarity among the two vectors.
The average cosine similarity results are reported in Table \ref{tab:similarity}.
The similarity between problems with the identified number in Table \ref{tab:similarity} and problem one implemented in Java, C++, and Ruby on average are $0.897$, $0.741$, and $0.564$, respectively. 
It is worth noting that this similarity analysis only focuses on tokens of each code snippet. 

The comparison of the experiment results in Tables \ref{tab:unseenproblems-rq2-3-java}, \ref{tab:unseenproblems-bccd-cplusplus}, and \ref{tab:unseen_problems_bccd_ruby} reveal that the base learners perform significantly better on Java compared to C++ and Ruby. 
The results in Table \ref{tab:similarity} reveal a discrepancy in the similarity between problem number one and the other problems for each programming language. 
The similarity scores show that the problems in the test set have the highest similarity with problem one (from the training set) in Java language, while this similarity is pretty low in Ruby.
This dissimilarity is reflected in the variation of F1-Scores among different problem sets. 
We believe this could be the reason that the models have higher scores for Java in the unseen problems scenario, compared to C++ or Ruby. 

However, it is noteworthy that the average F1-score in Ruby is slightly worse than that in C++, despite the fact that the problems in C++ are much more similar in comparison to Ruby. These results align with the previous findings in TBCCD \cite{yu2019neural}. 
One possible explanation is that apart from tokens, the structural similarity has an impact on the performance. Furthermore, it is important to note that the similarity between problem one and all other problems in all sets is not being calculated.

Another point to note in the results reported in Tables \ref{tab:unseenproblems-rq2-3-java}, \ref{tab:unseenproblems-bccd-cplusplus}, and \ref{tab:unseen_problems_bccd_ruby} is that TBCCD outperforms other baselines in almost all cases. TBCCD is particularly developed to address code clone detection. This model encodes the AST structure of code fragments using a convolutional approach along with the tokens of code snippets, which could contribute to its performance as it can compare code snippets more accurately. 

The final topic to address in this section is the underwhelming performance of ContraCode as the base learner in MAML. One possible explanation could be the approach to train the ContraCode. Since only JavaScript samples are used to generate positive and negative samples, it seems that the model cannot perform well for other programming languages. For example, we fine-tuned ContraCode on BigCloneBench and POJ-104 datasets and the results are $92.7$ and $43$, respectively; showing that it has a good performance for clone detection in Java, but not for Ruby.

\section{Scenario II: Unseen Programming Languages}\label{sec:scenario-ii}

In this section, we will discuss the approach, experimental setup, and the results for unseen languages.

\subsection{Approach}
The bottom part of Figure \ref{fig:training} shows the training process for this scenario. For evaluating the CCD models for different programming languages, we employed the same dataset of scenario I from CodeNet, where there are $105$ problems and their solutions are available in different languages. 
We made this choice because using the same problem sets across all languages enables the evaluation of unseen languages and minimizes the impact of unfamiliar problems. 
Here, we trained the models using the Java submissions of the dataset, as it is a widely studied language for CCD. For unseen languages, C++ and Ruby are considered to evaluate the performance of the models. Additionally, we repeated the same experiments by training models on C++ and evaluating them on Java and Ruby. After training the models on the source programming language, $15$ samples from the submission in each of the other programming languages are used to fine-tune the model. Finally, models are evaluated using the submissions of the selected unseen language. The fine-tuning and testing on each of the languages are done separately.\\

\textcolor{black}{Note that we only consider Java and C++ for training and then test on C++, Java, or Ruby as unseen language. We decided not to use Ruby as the source and the other two languages as unseen language in this scenario due to the following reasons. First, Ruby is the low resource language, and most of the CCD studies are conducted on Java and C++. Therefore, it is more reasonable to consider Ruby as an unseen language, which is closer to real world cases (i.e., we have a CCD model for Java or C++ and we want to adapt it for Ruby, not vice versa). Additionally, Ruby had lower scores in our previous experiments (scenario I). As training MAML is computationally expensive, it is also more reasonable to exclude Ruby here. Therefore, we conduct the following four cases: 1) Train on Java, test on C++ (shown as Java/C++); 2) train on Java, test on Ruby (Java/Ruby); 3) train on C++, test on Ruby (C++/Ruby); and 4) train on C++, test on Java (C++/Java). }

\subsection{Experimental Setup }

We used the same baselines, with the same hyperparameters, for unseen languages. As a result, For CodeBERT and RoBERTa, the learning rate remains $5e-5$. Moreover, the L2 norm is used to avoid overfitting. The number of episodes is also kept the same as unseen problems. 
We choose the learning rate to be $0.0002$ for TBCCD following our experiments for unseen problems.
We also keep $alpha$ and $beta$ for MAML unchanged. Since the downstream language is ignored in meta-learning, all episodes are formed using the base language. For instance, in the first experiment/line reported in Table \ref{tab:unseen-languages}, all episodes consist of Java snippets. In each episode, both clone and non-clone pairs exist. As a result, all episodes mimic a two-way 15-shot problem in the same language. After applying MAML on pre-trained models, we used ContraCode as a contrastive base learner. To make our experiments consistent with unseen problems, again, we used the same hyper-parameters for ContraCode, so we set the learning rate as $8e-4$. The Adam algorithm is used in the fine-tuning steps with $beta1 = 0.9$, $beta2=0.98$, and $epsilon=1e-6$, and the model is trained for $100$ epochs and $20,000$ gradient updates.

\subsection{Results}\label{resultsunseenlanguages}

In this section, the results of our experiments for binary code clone detection for unseen languages are reported.

\begin{table}[]
    \centering
    \caption{The results of the models for few-shot set up for binary code clone detection and when MAML is applied for \textbf{unseen languages}, the \textit{second scenario}. The F1-Scores are reported when the models are tested on sets of unseen languages.
    The first number is the results of CodeBERT, RoBERTa, and ContraCode in a $15$-shot learning setup for unseen languages, followed by the results of applying MAML with different base learners. The numbers in the two experiments are separated by ``/''. The number in the parentheses shows the improvement amount. In each experiment, the model is trained on the left language and tested on the right language. For example, Java/C++ means the model that is trained on Java is tested on C++.}
    
    \resizebox{0.95\columnwidth}{!}{%
    \begin{tabular}{@{}lcccc@{}}\toprule
   Language & CodeBERT& RoBERTA &ContraCode\\
    \midrule
        Java/C++ & 59.4/61.4 (\textbf{2}) &55.3/58.6 
        (\textbf{3.3}) & 57.1\\
        
        Java/Ruby & 37.5/39.6 (\textbf{2.1}) & 32.4/34.2 (\textbf{1.8}) & 33.7\\ 
        
        C++/ Ruby &35.4/37.1
        (\textbf{1.7}) & 29.1/30.3 (\textbf{1.2}) & 29.8\\
        
        C++/Java &57.4/59.2(\textbf{1.8}) & 52.2/55.1 (\textbf{2.9})
        & 54.3\\

    \bottomrule
        Mean & 47.42/49.32 (\textbf{1.9}) & 42.25/44.55 (\textbf{2.3}) & 43.72
    \end{tabular}}
    
    \label{tab:unseen-languages}
\end{table}

\subsubsection{RQ1: Performance of Models in Few-Shot Setup}

Table \ref{tab:unseen-languages} contains the results for the second scenario. 
In each row, there are two numbers that are separated by ``/'' and there is another number reported in the parenthesis. 
The first number shows the results of our experiments for different baselines for RQ1. 
In each row, we also show the training and test languages. For example, Java/C++ means the model is trained on Java and tested on C++ in a few-shot setting. 
As shown in Table \ref{tab:unseen-languages}, the F1-score is $59.4$ when CodeBERT is trained on Java, and C++ is employed as the downstream language. 
A similar score ($57.4$) is obtained from CodeBERT when we switch the place of Java and C++. 
When Ruby is used as the downstream programming language, CodeBERT achieves $37.5$ and $35.4$ for fine-tuning by Java and C++, respectively. \\
On average, CodeBERT results are $\sim5$ F1-scores higher than the ones from RoBERTa. In both models, when Ruby is the downstream language, the scores are in low-mid $30$s, but the models have higher scores if Java or C++ are the downstream languages.

\begin{figure*}
    \centering
    \includegraphics[width=1.1\textwidth]{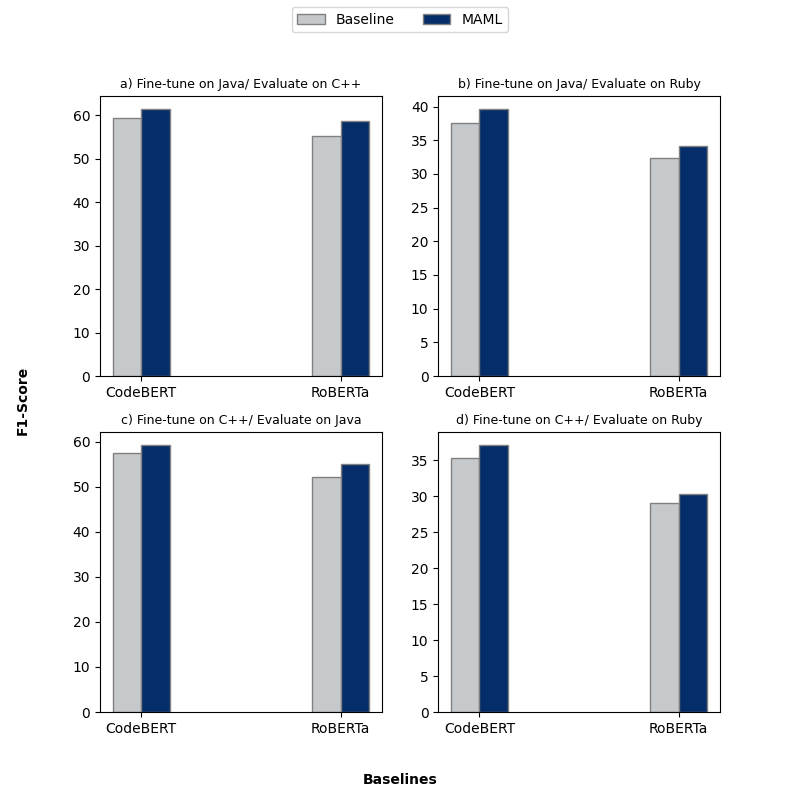}
    \caption{The results of CodeBERT and RoBERTa with and without MAML for binary classification in unseen languages: (a) Model trained using Java, evaluated using C++, (b) Model trained on Java, evaluated using Ruby, (c) Model trained using C++, evaluated on Java, (d) Model trained using C++ and evaluated on Ruby.} 
    \label{fig:unseen_languages_compare}
\end{figure*}

\subsubsection{RQ2: Performance of Models with Few-Shot Algorithm}

The second number after ``/" in Table \ref{tab:unseen-languages} shows the results of applying MAML to CodeBERT and RoBERTa. The number in the parenthesis is the F1-score improvement when MAML is used. 
On average, MAML improves the performance of CodeBERT by $1.9$, achieving an average F1-score of $49.32$. 
MAML applied on RoBERTa achieves an F1-score of $44.55$ on average, which shows an improvement of $2.3$ on average. 

Similar observations are seen here. When Ruby is the downstream language, the scores are much lower, compared to Java or C++ as the downstream language. The best results are when Java is used for training and C++ is the downstream language.
Figure \ref{fig:unseen_languages_compare} compares the results of baselines (\textcolor{gray}{gray} plot) with the performance of MAML (\textcolor{Blue}{blue} plot) applied to them. Though MAML improves the results, the improvement is not significant in this scenario. 


\subsubsection{RQ3: Performance of Models with Contrastive Learning}

The last column in Table \ref{tab:unseen-languages} shows the results of applying ContraCode as the base learner of MAML. The average score for ContraCode is $43.72$, and similar to previous RQs, the lower scores belong to Ruby in the test set. 
Compared to the results with CodeBERT/RoBERTa with/without MAML, ContraCode does not achieve better results and is between the scores for RoBERTa without and with MAML. 
This is aligned with our previous findings for unseen problems, where ContraCode marginally improved the results.

\begin{framed}
\textcolor{purple}{\textbf{Summary of Scenario II for Unseen Programming Languages in Binary CCD}} \\
The results of experiments for unseen languages for binary code clone detection confirm the vulnerability of these models for unseen languages. MAML and ContraCode with MAML improved the performance marginally.

\end{framed}

\subsection{Discussions}

\begin{figure*}
    \centering
    \includegraphics[width=1.1\textwidth]{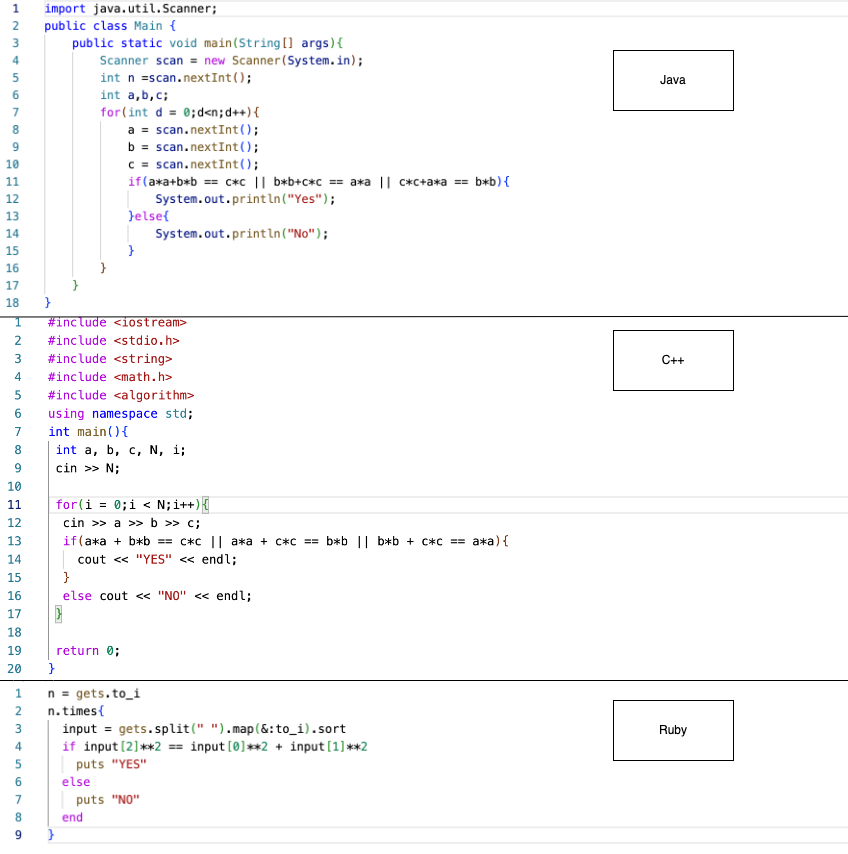}
    \caption{Three sample submissions for problem \#3 in our dataset which are implemented in Java (top), C++ (middle), and Ruby (bottom). The code snippets are providing solutions for a problem asking with three integers, the code should indicate if there exists a triangle. The samples show the similarity of the tokens and code structure between Java and C++, and how they are different from Ruby (yet, there are some similarities between Ruby and other programming languages).} 
    \label{fig:codenetproblem3}
\end{figure*}

We observed that when Ruby is the downstream language, the performance of the models is much lower than cases when Java or C++ are the downstream languages. We relate this to the dis/similarity among the languages.
Figure \ref{fig:codenetproblem3} is an example of submissions for Problem \#$3$ in the CodeNet dataset, for which, the code should detect if three integers can form a triangle, and it shows the solutions in Java, C++, and Ruby.
As shown in Figure \ref{fig:codenetproblem3}, the code written in Java and C++ have similar tokens and structures but are different from Ruby.
Using the same approach as described previously in Section \ref{subsec:scenario1-discussions}, the Java and C++ codes have a cosine similarity of $0.91$. On the other hand, the similarity between Java and C++ with Ruby are $0.64$ and $0.70$, respectively. 
As a result, a model trained in Java can perform well on CCD in C++ and vice versa. But, when the solution is implemented in Ruby, it has slightly different tokens and structure; thus has lower scores.

In the case of unseen languages, although the Ruby implementations have some similarities, the score is lower compared to Java and C++. However, the F1-score is still higher than the results for unseen problems. 
As Table \ref{tab:unseen-languages} shows, the overall scores of baselines without MAML are higher than the results achieved for unseen problems. 
We relate this to the fact that although the implementations are different in various languages, as they relate to similar problems, they still share some semantic and lexical similarities. However, in the case of unseen problems, not only the differences in the languages happen, the new problems have new sets of tokens that can affect how the model learns about the vocabulary and henceforth, degrade the results.

Table \ref{tab:unseen-languages} shows that MAML can improve the F1-score of CodeBERT by $1.9$ and the F1-score of RoBERTa by $2.3$. Yet, compared to the results reported for unseen problems, this improvement is negligible. The main reason for this lack of increase in MAML's performance can be traced back to episodic-based training. In the case of unseen problems, each episode consists of different problems, so the meta-learner is trained on tasks that are relevant to the downstream tasks. As a result, in the case of unseen problems, each problem set is treated as a new class. The meta-learner is trained on many unseen problems in the meta-learning process. The model then aims to solve the clone detection task for an unseen problem that was not seen during meta-learning. On the other hand, in the case of unseen languages, the meta-learner is trained on code snippets of the same language, so it does not learn to solve code clone detection in different languages.

\section{Scenario III: Unseen Problems and Unseen Languages} \label{sec:scenario-iii}

In this section, we evaluate the performance of models for the third scenario. We first discuss the approach and then provide the results of the classification task, followed by the results of the code retrieval clone detection.

\subsection{Approach and Experimental Setup}

The third scenario is a combination of the previous two cases, scenario I, which pertains to unknown problems, and Scenario II, which focuses on unseen languages. 
For this scenario, we first fine-tuned the baselines using the first $15$ problems of the source language (e.g., Java).
Then, we exposed models to $15$ samples of each distinct problem from various problem sets of the downstream language (i.e., unseen language, such as C++). Finally, we evaluated the models against various problem sets of the downstream language (e.g., C++). The same experiments are applied for both binary code clone detection and retrieval-based code clone detection.
For binary classification, we use the CodeNet\textsubscript{B} and for the retrieval-based task, we use the CodeNet\textsubscript{R} dataset. 
Additionally, as CodeBERT outperformed RoBERTa in our experiments, we only consider CodeBERT in this section. 

To ensure comparability, we keep all of the hyper-parameters for the baselines consistent with those used in the previous scenarios, as outlined in Section \ref{sec:scenario-i}. 
The learning rate for CodeBERT is set to $5e-5$, and l2-norm is used to prevent overfitting. 
TBCCD is designed for binary classification, so TBCCD is not considered in retrieval-based experiments. Yet, the hyper-parameters for TBCCD for the third scenario remain the same as in previous sections. 
 We set $\alpha$ and $\beta$ in MAML to $1e-5$. The same hyper-parameters as previously described are used for ContraCode (i.e., the learning rate is $8e-4$, Adam is the optimizer algorithm with $\beta_{1} = 0.9$, $\beta_{2} = 0.98$, and $\epsilon = 1e-6$). 

\textcolor{black}{As training the models is computationally expensive, and based on the observed results in previous sections, we made some adjustments to our experiments. First, we eliminated RoBERTa, as in previous scenarios, CodeBERT achieved higher results. Therefore, we only consider CodeBERT and TBCCD in the third scenario for binary CCD to answer RQ1 and RQ2. RQ3 only uses ContraCode which remains the same here. Additionally, we only apply retrieval-based CCD in the third scenario. A main factor in this decision was the extensive time it takes to train each of the models and based on the observed results, which were low in all cases. }

\subsection{Results}

We present the results of the third scenario in this section. For each RQ, we report the obtained scores for both binary classification and retrieval-based clone detection tasks. 

\begin{table}[]
    \centering
    \caption{The results of the models for few-shot setup and when MAML is applied for (source) Java and (unseen) C++ in the \textit{third scenario}, for \textbf{binary classification CCD}. The F1-Scores are reported when the models are tested on sets of unseen problems and unseen languages. The first number is the results of CodeBERT and TBCCD in 15-shot learning setup for unseen problems (RQ1), followed by the results of applying MAML with different base learners (RQ2). The numbers in the two experiments are separated by “/”. The number in the parentheses shows the improvement score. The last column are the results when ContraCode is used in MAML (RQ3).}
    \resizebox{0.95\columnwidth}{!}{%
    \begin{tabular}{@{}lccc@{}}\toprule
   Problem \# & CodeBERT& TBCCD & ContraCode\\
    \midrule

        P \#16-30 & 25.4/46.3(\textbf{20.9}) & 25.1/47.3(\textbf{22.2}) & 47.6\\
        
        P \#31-45 & 24.8/45.6(\textbf{20.8}) & 24.4/46.9(\textbf{22.5}) & 48.8\\ 
        
        P \#46-60 & 24.7/45.9(\textbf{21.2}) & 24.7/46.5(\textbf{21.8})& 48.1\\
    
        P \#61-75 &25.1/46.7(\textbf{21.6}) & 23.9/46.6(\textbf{22.7}) & 47.3\\
        
        P \#76-90 & 24.1/45.8(\textbf{21.7}) & 23.8/45.8(\textbf{22})& 48.2\\
        
        P \#91-105 & 25.5/46.9(\textbf{21.4}) & 24.7/46.9(\textbf{22.2})& 47.2\\

    \bottomrule
    Mean & 24.93/46.2 (\textbf{21.27}) & 24.43/46.66 (\textbf{22.23}) & 47.86
    \end{tabular}}
    
    \label{tab:ThirdScenarion:JavaCPlusPlusbinary}
\end{table}

\begin{table}[]
    \centering
    \caption{The results of the models for few-shot setup and when MAML is applied for (source) Java and (unseen) Ruby in the \textit{third scenario}, for \textbf{binary classification CCD}. The F1-Scores are reported when the models are tested on sets of unseen problems and unseen languages. The first number is the results of CodeBERT and TBCCD in a 15-shot learning setup for unseen problems (RQ1), followed by the results of applying MAML with different base learners (RQ2). The numbers in the two experiments are separated by “/”. The number in the parentheses shows the improvement score. The last column shows the results when ContraCode is used as a base learner in MAML (RQ3).}
    \resizebox{0.95\columnwidth}{!}{%
    \begin{tabular}{@{}lccc@{}}\toprule
   Problem \# & CodeBERT& TBCCD & ContraCode\\
    \midrule

        P \#16-30 & 20.2/39.5 (\textbf{19.3}) & 21.3/40.2(\textbf{18.9}) & 41.2\\
        
        P \#31-45 & 20.3/40.1 (\textbf{19.8}) & 20.2/39.7(\textbf{19.5}) & 40.3\\ 
        
        P \#46-60 & 19.5/39.3(\textbf{19.8}) & 20.7/40.7(\textbf{20})& 41.8\\
    
        P \#61-75 &21.2/40.3(\textbf{19.1}) & 19.8/39.5(\textbf{19.7}) & 39.9\\
        
        P \#76-90 & 20.5/39.8(\textbf{19.3}) & 21.4/40.6 (\textbf{19.2})& 41.7\\
        
        P \#91-105 & 19.7/38.8(\textbf{19.1}) & 20.3/40.1(\textbf{19.8})& 40.2\\

    \bottomrule
    Mean & 20.23/39.63 (\textbf{19.4}) & 20.61/40.13 (\textbf{19.52}) & 40.85    
    \end{tabular}}
    
    \label{tab:thirdScenarion-JavaRuby}
\end{table}

\begin{table}[]
    \centering
    \caption{The results of the models for few-shot setup and when MAML is applied for (source) C++ and (unseen) Ruby in the \textit{third scenario}, for \textbf{binary classification CCD}. The F1-Scores are reported when the models are tested on sets of unseen problems and unseen languages. The first number is the results of CodeBERT and TBCCD in a 15-shot learning setup for unseen problems (Rq1), followed by the results of applying MAML with different base learners (RQ2). The numbers in the two experiments are separated by “/”. The number in the parentheses shows the improvement score. The last column is the obtained scores when ContraCode is used in MAML (RQ3).}
    \resizebox{0.95\columnwidth}{!}{%
    \begin{tabular}{@{}lccc@{}}\toprule
   Problem \# & CodeBERT& TBCCD & ContraCode\\
    \midrule

        P \#16-30 & 18.9/39.1 (\textbf{20.2}) & 19.8/40.5 (\textbf{20.7})& 40.3\\
        
        P \#31-45 & 19.2/39.6 (\textbf{20.4})& 20.2/40.8 (\textbf{20.6})& 41.4\\ 
        
        P \#46-60 & 19.8/ 40.1 (\textbf{20.3})& 20.1/41.1 (\textbf{21})& 40.9\\
    
        P \#61-75 & 20.1/ 40.4 (\textbf{20.3})& 20.5/ 41.3 (\textbf{20.8})& 41.2\\
        
        P \#76-90 & 19.6/ 39.3 (\textbf{19.7}) & 20.4/ 40.7 (\textbf{20.3})& 41.2\\
        
        P \#91-105 & 19.8/ 39.4 (\textbf{19.6}) & 20.7/ 40.8 (\textbf{20.1})& 41.4\\
   \bottomrule
   Mean & 19.56/39.65 (\textbf{20.09}) & 20.28/40.86 (\textbf{20.58}) & 41.06
    \end{tabular}}
    
    \label{tab:thirdScenarion-cplusplusRubybinary}
\end{table}

\subsubsection{RQ1: Performance of Models in Few-Shot Setup}

The results for the third scenario for binary classification CCD are presented in Tables \ref{tab:ThirdScenarion:JavaCPlusPlusbinary} to \ref{tab:thirdScenarion-cplusplusRubybinary}. 
The first numbers before ``/" represent the results for a few-shot setting of CodeBERT and TBCCD for the third scenario. 
Interestingly, the results obtained from CodeBERT and TBCCD are similar, regardless of the source and downstream language. The scores are around $25$ when Java is the source and the downstream language is C++, which are improved to $\sim 46$ F1-score when MAML is used. 
Similar to the second scenario, when Ruby is the unseen language, the scores are lower, $\sim 20$ without MAML and $\sim 40$ when MAML is used, both for CodeBERT and \textcolor{purple}{TBCCD}, regardless of the source language being Java or C++.

The MAP@R scores for the retrieval-based code clone detection are shown in Tables \ref{tab:ThirdScenario:JavaCPlusPlus-CR} to \ref{tab:ThirdScenarion-coderetrieval-cppruby} (see the first number before ``/").
The results for the retrieval-based CCD are lower than the ones we observed for the binary classification, and they range from $\sim 13$ to $\sim 18$ when we apply CodeBERT in the few-shot setting. 
Similar to our previous observations, when Ruby is the unseen language, the scores are the lowest.

\begin{table}[]
    \centering
    \caption{The results of the models for few-shot setup and when MAML is applied for (source) Java and (unseen) C++ in the \textit{third scenario} for \textbf{Code Retrieval Task}. The MAP@R is reported when the models are tested on sets of unseen problems and unseen languages. The first number is the results of CodeBERT in a 15-shot learning setup (RQ1), followed by the results of applying MAML with different base learners (RQ2). The numbers in the two experiments are separated by “/”. The number in the parentheses shows the improvement score. The last column is when ContraCode is used as MAML's base learner (RQ3).}
    \resizebox{0.69\columnwidth}{!}{%
    \begin{tabular}{@{}lccc@{}}\toprule
   Problem \# & CodeBERT & ContraCode\\
    \midrule

        P \#16-30 & 18.2/41.7 (\textbf{23.5})  & 41.2\\
        
        P \#31-45 & 17.9/39.6 (\textbf{21.7}) & 39.1\\ 
        
        P \#46-60 & 17.1/41.2 (\textbf{22.1}) & 41.9\\
    
        P \#61-75 & 17.7/38.5 (\textbf{20.8}) & 39.1\\
        
        P \#76-90 & 18.1/40.9 (\textbf{22.8}) & 40.4\\
        
        P \#91-105 & 19.3/41.8 (\textbf{22.5}) & 41.1\\

    \bottomrule
    Mean &  18.38/40.61 \textbf{(22.23)}& 40.46
    \end{tabular}}
    
    \label{tab:ThirdScenario:JavaCPlusPlus-CR}
\end{table}

\begin{table}[]
    \centering
    \caption{The results of the models for few-shot setup and when MAML is applied for (source) Java and (unseen) Ruby in the \textit{third scenario} for \textbf{Code Retrieval Task}. The MAP@R is reported when the models are tested on sets of unseen problems and unseen languages. The first number is the results of CodeBERT in a 15-shot learning setup (RQ1), followed by the results of applying MAML with different base learners (RQ2). The numbers in the two experiments are separated by “/”. The number in the parentheses shows the improvement score. The last column is when ContraCode is used as MAML's base learner (RQ3).}
    \resizebox{0.69\columnwidth}{!}{%
    \begin{tabular}{@{}lccc@{}}\toprule
   Problem \# & CodeBERT & ContraCode\\
    \midrule

        P \#16-30 & 14.5/38.7 (\textbf{24.2})  & 39.4\\
        
        P \#31-45 & 15.3/38.6 (\textbf{23.3}) & 37.6\\ 
        
        P \#46-60 & 12.4/38.2 (\textbf{25.8}) & 36.3\\
    
        P \#61-75 & 14.9/39.5 (\textbf{24.6}) & 39.9\\
        
        P \#76-90 & 12.9/38.9 (\textbf{26}) & 36.9\\
        
        P \#91-105 & 14.8/40.1 (\textbf{25.3}) & 36.5\\

    \bottomrule
    Mean & 14.13/39.00 (\textbf{24.86})& 37.76
    \end{tabular}}
    
    \label{tab:ThirdScenarion-coderetrieval-javaruby}
\end{table}

\begin{table}[]
    \centering
    \caption{The results of the models for few-shot setup and when MAML is applied for (source) C++ and (unseen) Ruby in the \textit{third scenario} for \textbf{Code Retrieval Task}. The MAP@R is reported when the models are tested on sets of unseen problems and unseen languages. The first number is the results of CodeBERT in a 15-shot learning setup (RQ1), followed by the results of applying MAML with different base learners (RQ2). The numbers in the two experiments are separated by “/”. The number in the parentheses shows the improvement score. The last column is when ContraCode is used as MAML's base learner (RQ3).}
    \resizebox{0.69\columnwidth}{!}{%
    \begin{tabular}{@{}lccc@{}}\toprule
   Problem \# & CodeBERT & ContraCode\\
    \midrule

        P \#16-30 & 12.5/35.3 (\textbf{22.8})  & 36.4\\
        
        P \#31-45 & 13.3/36.7 (\textbf{23.4}) & 37.1\\ 
        
        P \#46-60 & 13.4/36.9 (\textbf{23.5}) & 36.8\\
    
        P \#61-75 & 13.9/37.3 (\textbf{23.4}) & 35.1\\
        
        P \#76-90 & 11.9/35.5 (\textbf{23.6}) & 37.2\\
        
        P \#91-105 & 12.7/37.2 (\textbf{24.5}) & 35.9\\

    \bottomrule
    Mean & 12.85/36.48 (\textbf{23.53})& 36.25
    \end{tabular}}
    
    \label{tab:ThirdScenarion-coderetrieval-cppruby}
\end{table}

\begin{figure*}
    \centering
    \includegraphics[width=1.1\textwidth]{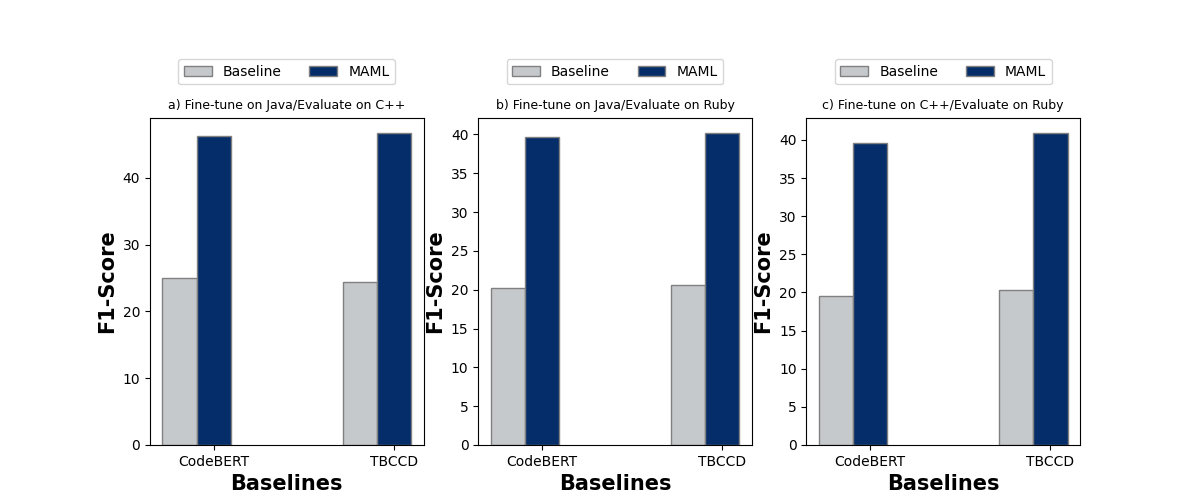}
    \caption{Comparison of the F1-scores for CodeBERT and TBCCD, without (\textcolor{gray}{gray} plots) and with (\textcolor{Blue}{blue} plots) MAML in a few-shot setting for the third scenario in binary code clone detection. The left plot is when the models are fine-tuned on Java and evaluated on C++, the middle plot shows the plots for models that are fine-tuned on Java and evaluated on Ruby, and the right plot is when the models are fine-tuned on C++ and evaluated on Ruby.} 
    \label{fig:compare-third-binary}
\end{figure*} 

\subsubsection{RQ2: Performance of CodeBERT with Few-Shot Algorithm}

The second numbers reported in Tables \ref{tab:ThirdScenarion:JavaCPlusPlusbinary} to \ref{tab:thirdScenarion-cplusplusRubybinary} and Tables \ref{tab:ThirdScenario:JavaCPlusPlus-CR} to \ref{tab:ThirdScenarion-coderetrieval-cppruby} (after ``/") indicate the results when MAML is used, for binary classification and retrieval-based clone detection. 
In both tasks, MAML helps improve the result by $\sim 20$ - $\sim 25$ F1/MAP@R scores, boosting them to $\sim 40$. The score gain for the retrieval-based approach is a bit more when MAML is used. 

\begin{figure*}
    \centering
    \includegraphics[width=1.1\textwidth]{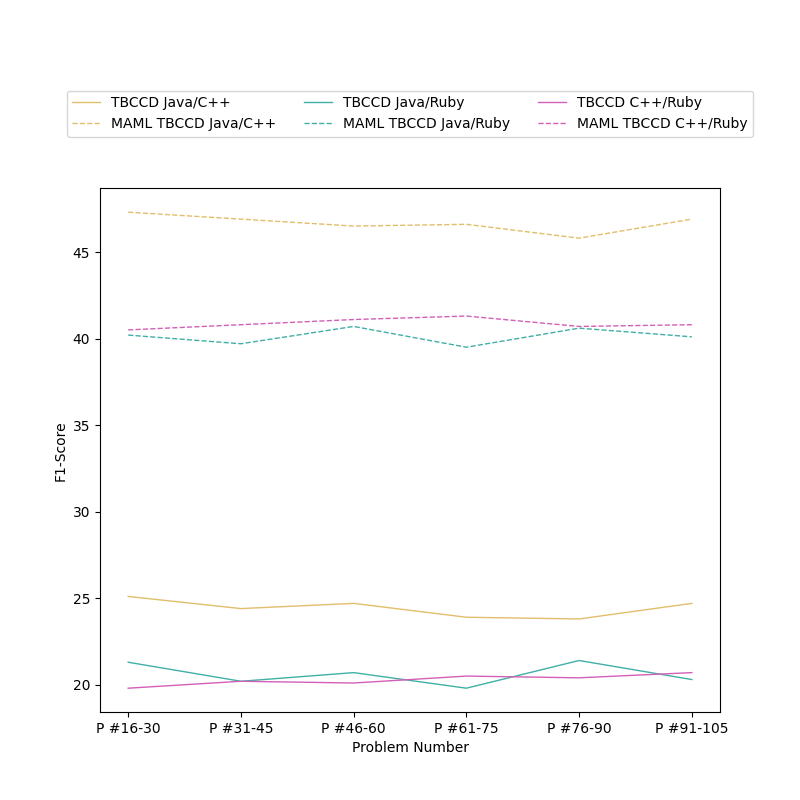}
    \caption{The F1-score comparison of TBCCD without (solid lines) and with (dashed lines) MAML for the third scenario in the binary code clone detection across different problem sets. The train/test languages are shown beside the model name. For example, TBCCD Java/Ruby shows the results of TBCCD when it is trained on Java and tested on Ruby (an unseen language). The colors represent the results of TBCCD with different train/test languages. Using MAML helps in increasing the performance by $\sim 20$ scores.}
    \label{fig:unseen_third_binary}
\end{figure*}

\begin{figure*}
    \centering
    \includegraphics[width=1.1\textwidth]{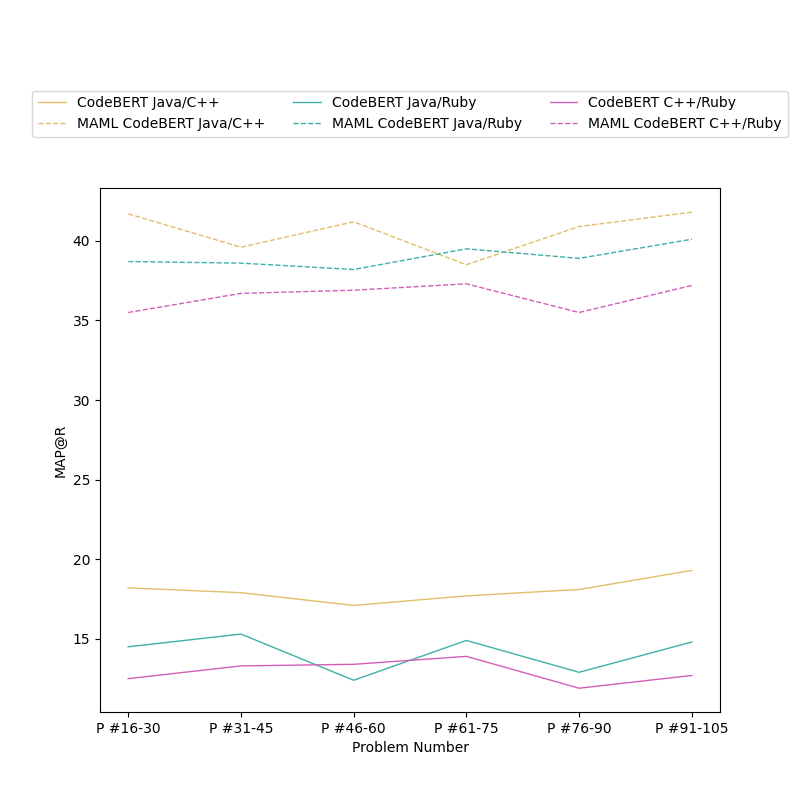}
    \caption{MAP@R score of CodeBERT without (solid lines) and with (dashed lines) MAML for the third scenario, for the retrieval-based clone detection task. The two languages used for fine-tuning and testing are separated with `/' as train/test. For example, \textit{MAML CodeBERT Java/C++} shows the results of MAML with CodeBERT when the model is trained on Java and is tested on C++ (the unseen language). The scores are shown across different problem sets, and the colors show models with various training/testing languages. Using MAML boosts the performance by $\sim 20$ scores or a bit more.}
    \label{fig:unseen_third_retrieval}
\end{figure*}

Figure \ref{fig:compare-third-binary} compares the results of baselines (\textcolor{gray}{gray} plots) in a few-shot setting with the results when MAML is applied (\textcolor{Blue}{blue} plots). 
Figures \ref{fig:unseen_third_binary} and \ref{fig:unseen_third_retrieval} have close-up plots for the results of TBCCD and CodeBERT for RQ1 and RQ2, respectively, for binary classification and retrieval-based tasks.
In these two figures, the scores are shown for each set of problems, without (solid lines) and with (dashed lines) MAML, and the settings for different source and target languages are shown with colors. Though there are some ups and downs in the scores for each problem set, overall, there is not much difference among the scores for each group of problems. But the differentiation among the scores for different source and unseen test languages is bolder, where higher scores are achieved when the model is trained on Java and tested on C++ (shown as Java/C++). 

\subsubsection{RQ3: Performance of Models with Contrastive Learning}

The last column of Tables \ref{tab:ThirdScenarion:JavaCPlusPlusbinary} to \ref{tab:thirdScenarion-cplusplusRubybinary} (binary classification) and Tables \ref{tab:ThirdScenario:JavaCPlusPlus-CR} to \ref{tab:ThirdScenarion-coderetrieval-cppruby} (retrieval-based CCD) represent the results when ContraCode is integrated with MAML. 
Using contrastive objectives in MAML improves the results compared to when CodeBERT is MAML's base learner. However, similar to the previous scenarios, the performance boost is negligible. 
Even in the case of retrieval tasks when Ruby is the unseen language, the results are degraded slightly.

\begin{framed}
    The results are similar to previous scenarios. MAML can boost the performance of the models significantly, and, although ContraCode slightly upgrades the scores, the increase in the scores is small. 
\end{framed}

\subsection{Discussions}

For the third scenario, we explored both the binary classification and retrieval-based tasks of clone detection in the few-shot setting. 
Interestingly, the retrieval-based results are lower than the binary classification, and it might indicate that the retrieval task is a harder one, in terms of prediction scores. 

Similar to previous sections, the models' scores in the few-shot setting drop significantly compared to fully fine-tuning the models, showing that the models cannot perform in the few-shot setting. In both tasks, when MAML is applied, the performance increased significantly. 
However, the scores were still below $50$. 
Even using a contrastive objective was not helpful, and similar to our previous experiments in scenarios I and II, the improvement is negligible. The poor performance of ContraCode is likely due to its pre-training language which is JavaScript. 
Additionally, we observed small variations in the scores of models (see Figures \ref{fig:unseen_third_binary} and \ref{fig:unseen_third_retrieval}) across different problem sets. These inconsistencies align with the results from previous sections, as we noted that certain problem sets contain more similar tokens, resulting in varying performances when evaluating each set.

\section{Implications} \label{sec:implications}

Our findings highlight the limitations of deep learning and pre-trained language models for code clone detection in a few-shot setting. Despite the promising results of few-shot algorithms like MAML, there remains a performance gap between fully fine-tuned models and those trained using few-shot learning. Specifically, MAML is unable to improve baseline performance for unseen programming languages, which is a major challenge when constructing episodes that facilitate learning across multiple languages.
Addressing this problem or developing other few-shot approaches is a future line of research. 

Additionally, we found that contrastive learning was not helpful in improving the results, beyond what MAML already provided. 
To address this problem, one possible solution is to adopt a training approach that leverages knowledge from diverse programming languages. One approach involves creating a training set that combines two languages, such as Java and C++, and training a meta-learner on this mixed-language dataset. By including multiple languages in the training process, this approach might enhance the contrastive learning aspect of the study.
Furthermore, the current version of the model used to address RQ3 is pre-trained with Javascript. Thus, fine-tuning this model with different programming languages may improve the performance of this contrastive learning method.

Overall, the limitations of current few-shot algorithms for code clone detection in unfamiliar languages suggest that further research is needed to develop more robust models. Adopting a training approach that incorporates knowledge from diverse programming languages may be a promising direction for future work in this area. Ultimately, improving the performance of code clone detection models in few-shot settings could have significant practical implications, such as enabling more efficient and accurate code reuse across programming languages.

Researchers could also work on developing algorithms that are specifically designed to capture the needs of source code. Although programming and natural languages have similarities, there are structural and other differences among them, and algorithms that consider these differences might be more helpful. Interestingly, MAML is another field is known to help the performance of the models with certain limits (usually does not go beyond $60-70\%$) \cite{walsh2022automated}, which brings the necessity of developing other algorithms for few-shot setting. 

\section{Threats to Validity}\label{sec:threats} 

\textbf{Internal Validity}
Internal validity is associated with having unwanted results. 
To mitigate this threat, we used publicly available datasets and followed a similar approach to prepare the CCD datasets with a similar schema as the benchmark datasets. 
The CodeNet dataset is a high-quality one and is used in other studies \cite{huang2021codenet}. 
We also used common evaluation metrics employed by previous studies to report the results. Another threat in this category can be related to running the models. To reduce the effects of such threat, we followed previous works to choose hyperparameters for the few-shot learning algorithm, and we also employed the reported hyperparameters of the baseline models. 
All the experiments are executed on the same machine with the same configuration to make it easy to reproduce the results. So, we anticipate a low internal threat in our work.

\textbf{External Validity}
External threats in our work related to the generalizability of the results to other models and programming languages. Although the approach and the methodology do not differ for other programming languages and models, the results might not be generalizable to other programming languages.
We also only study the CCD in our work. So though the CLCCD and code clone search are closely related fields, generalizing the results to these two fields or other applications requires separate research.

\textbf{Construct Validity}
The metrics we chose for the evaluation of the models are the widely used and acceptable metrics for our tasks. Both F1-score and MAP@R are previously used and reported in other CCD works. So, we do not anticipate a threat related to choosing the evaluation metrics. 
Another threat could be related to choosing an appropriate few-shot learning algorithm. We conducted a literature review on few-shot learning algorithms and chose the algorithm that is widely used \cite{wang2021grad2task,li2021semi,bansal2022few} and referenced for different applications, including natural language processing. Although few-shot learning techniques have not been used in software engineering for CCD, MAML is used for natural language processing previously \cite{wang2021grad2task}, which is the closest application to code (compared to computer vision). Therefore, we do not anticipate a threat related to the choice of algorithm, not to mention that this is an empirical study to evaluate the application of few-shot for CCD. 

\textcolor{black}{Another threat could be related to constructing the dataset, where solutions for one problem are considered as clones of each other, and the solutions of other problems are considered as the negative class. In this setting, we chose the entire functions in the solution into account and did not check whether some parts of the implementations are similar across the solutions of various problems. This might bring a threat to the validity of the results. However, as the input and output of various problems are different, we anticipate such threat to be low, considering that we care about the full functionality of the code snippets, rather than finding their similarity percentage. }


\section{Conclusion and Future Works}\label{sec:conclusion}

In this paper, we investigated the performance of various models for code clone detection in the few-shot setting. The findings suggest that current state-of-the-art models underperform when evaluated in a few-shot setting. Furthermore, our experiments showed that using MAML, a well-known few-shot learning algorithm could improve the performance of code clone detection models when only a few samples are available. However, the obtained scores are still much below the scores when models use the full dataset; this improvement is more required in the case of unseen languages. 
Though contrastive learning is shown in previous studies to be helpful for source code representation, combining contrastive learning using ContraCode with MAML only boosts the performance slightly, and in some cases decreases the scores. 

The limitations of current models for code clone detection in the few-shot setting demonstrate the need for further research in developing more effective methods. Future research should focus on developing models that can better handle the challenge of detecting code clones in unfamiliar programming languages and unseen problem sets. Additionally, investigating alternative approaches to contrastive learning may provide new insights and opportunities for improving the performance of code clone detection models in the few-shot setting.
Overall, the research presented in this paper contributes to the ongoing efforts to develop robust models for code clone detection in the few-shot setting. 

\section{Data Availability Statement}

The data used in this study was obtained from CodeNet, a publicly available dataset of over $14$ million code samples from GitHub. CodeNet was created and made available by the Allen Institute for AI (AI2) and includes code samples in six programming languages: Python, Java, JavaScript, Ruby, Go, and PHP.
CodeNet can be accessed through the AI2 website \footnote{{https://github.com/IBM/Project\_CodeNet}} and is distributed under the Creative Commons Attribution 4.0 International (CC BY 4.0) license. The dataset is available for research and non-commercial use and can be downloaded free of charge. \textcolor{black}{Additionally, we have open-sourced our scripts to enable the research community to replicate this research}\footnote{{https://github.com/mkhfring/EMSE\_Replication.git}}.

\section{Conflict of Interest}
This research is supported by a grant from the Natural Sciences and Engineering Research Council of Canada RGPIN-2019-05175.

\newpage
\bibliographystyle{plain}
\bibliography{main}

\begin{thebibliography}{10}

\bibitem{ahmad-etal-2021-unified}
Wasi Ahmad, Saikat Chakraborty, Baishakhi Ray, and Kai-Wei Chang.
\newblock Unified pre-training for program understanding and generation.
\newblock In {\em Proceedings of the 2021 Conference of the North American
  Chapter of the Association for Computational Linguistics: Human Language
  Technologies}, pages 2655--2668, Online, June 2021. Association for
  Computational Linguistics.

\bibitem{ain2019systematic}
Qurat~Ul Ain, Wasi~Haider Butt, Muhammad~Waseem Anwar, Farooque Azam, and Bilal
  Maqbool.
\newblock A systematic review on code clone detection.
\newblock {\em IEEE access}, 7:86121--86144, 2019.

\bibitem{aitchison2021infonce}
Laurence Aitchison.
\newblock Infonce is a variational autoencoder.
\newblock {\em arXiv preprint arXiv:2107.02495}, 2021.

\bibitem{ankali2021detection}
Sanjay~B Ankali and Latha Parthiban.
\newblock Detection and classification of cross-language code clone types by
  filtering the nodes of antlr-generated parse tree.
\newblock {\em International Journal of Intelligent Systems and Applications},
  13(3):43--65, 2021.

\bibitem{bansal2022few}
TRAPIT BANSAL.
\newblock {\em FEW-SHOT NATURAL LANGUAGE PROCESSING BY META-LEARNING WITHOUT
  LABELED DATA}.
\newblock PhD thesis, University of Massachusetts Amherst, 2022.

\bibitem{bansal-etal-2020-learning}
Trapit Bansal, Rishikesh Jha, and Andrew McCallum.
\newblock Learning to few-shot learn across diverse natural language
  classification tasks.
\newblock In {\em Proceedings of the 28th International Conference on
  Computational Linguistics}, pages 5108--5123, Barcelona, Spain (Online),
  December 2020. International Committee on Computational Linguistics.

\bibitem{bareiss2022code}
Patrick Barei{\ss}, Beatriz Souza, Marcelo d'Amorim, and Michael Pradel.
\newblock Code generation tools (almost) for free? a study of few-shot,
  pre-trained language models on code.
\newblock {\em arXiv preprint arXiv:2206.01335}, 2022.

\bibitem{bellon2007comparison}
Stefan Bellon, Rainer Koschke, Giulio Antoniol, Jens Krinke, and Ettore Merlo.
\newblock Comparison and evaluation of clone detection tools.
\newblock {\em IEEE Transactions on software engineering}, 33(9):577--591,
  2007.

\bibitem{brown2020language}
Tom Brown, Benjamin Mann, Nick Ryder, Melanie Subbiah, Jared~D Kaplan, Prafulla
  Dhariwal, Arvind Neelakantan, Pranav Shyam, Girish Sastry, Amanda Askell,
  et~al.
\newblock Language models are few-shot learners.
\newblock {\em Advances in neural information processing systems},
  33:1877--1901, 2020.

\bibitem{chen2022transferability}
Fuxiang Chen, Fatemeh~H Fard, David Lo, and Timofey Bryksin.
\newblock On the transferability of pre-trained language models for
  low-resource programming languages.
\newblock In {\em Proceedings of the 30th IEEE/ACM International Conference on
  Program Comprehension}, pages 401--412, 2022.

\bibitem{chen2021evaluating}
Mark Chen, Jerry Tworek, Heewoo Jun, Qiming Yuan, Henrique Ponde de~Oliveira
  Pinto, Jared Kaplan, Harri Edwards, Yuri Burda, Nicholas Joseph, Greg
  Brockman, et~al.
\newblock Evaluating large language models trained on code.
\newblock {\em arXiv preprint arXiv:2107.03374}, 2021.

\bibitem{devlin2018bert}
Jacob Devlin, Ming-Wei Chang, Kenton Lee, and Kristina Toutanova.
\newblock Bert: Pre-training of deep bidirectional transformers for language
  understanding.
\newblock {\em arXiv preprint arXiv:1810.04805}, 2018.

\bibitem{feng-etal-2020-codebert}
Zhangyin Feng, Daya Guo, Duyu Tang, Nan Duan, Xiaocheng Feng, Ming Gong, Linjun
  Shou, Bing Qin, Ting Liu, Daxin Jiang, and Ming Zhou.
\newblock {C}ode{BERT}: A pre-trained model for programming and natural
  languages.
\newblock In {\em Findings of the Association for Computational Linguistics:
  EMNLP 2020}, pages 1536--1547, Online, November 2020. Association for
  Computational Linguistics.

\bibitem{finn2017model}
Chelsea Finn, Pieter Abbeel, and Sergey Levine.
\newblock Model-agnostic meta-learning for fast adaptation of deep networks.
\newblock In {\em International Conference on Machine Learning}, pages
  1126--1135. PMLR, 2017.

\bibitem{GuoRLFT0ZDSFTDC21}
Daya Guo, Shuo Ren, Shuai Lu, Zhangyin Feng, Duyu Tang, Shujie Liu, Long Zhou,
  Nan Duan, Alexey Svyatkovskiy, Shengyu Fu, Michele Tufano, Shao~Kun Deng,
  Colin~B. Clement, Dawn Drain, Neel Sundaresan, Jian Yin, Daxin Jiang, and
  Ming Zhou.
\newblock Graphcodebert: Pre-training code representations with data flow.
\newblock In {\em 9th International Conference on Learning Representations,
  {ICLR} 2021, Virtual Event, Austria, May 3-7, 2021}. OpenReview.net, 2021.

\bibitem{huang2021codenet}
Qijing Huang, Dequan Wang, Zhen Dong, Yizhao Gao, Yaohui Cai, Tian Li, Bichen
  Wu, Kurt Keutzer, and John Wawrzynek.
\newblock Codenet: Efficient deployment of input-adaptive object detection on
  embedded fpgas.
\newblock In {\em The 2021 ACM/SIGDA International Symposium on
  Field-Programmable Gate Arrays}, pages 206--216, 2021.

\bibitem{jain-etal-2021-contrastive}
Paras Jain, Ajay Jain, Tianjun Zhang, Pieter Abbeel, Joseph Gonzalez, and Ion
  Stoica.
\newblock Contrastive code representation learning.
\newblock In {\em Proceedings of the 2021 Conference on Empirical Methods in
  Natural Language Processing}, pages 5954--5971, Online and Punta Cana,
  Dominican Republic, November 2021. Association for Computational Linguistics.

\bibitem{Keivanloo2021}
Iman Keivanloo and Juergen Rilling.
\newblock {\em Source Code Clone Search}, pages 121--134.
\newblock Springer Singapore, Singapore, 2021.

\bibitem{lachaux2021dobf}
Marie-Anne Lachaux, Baptiste Roziere, Marc Szafraniec, and Guillaume Lample.
\newblock Dobf: A deobfuscation pre-training objective for programming
  languages.
\newblock {\em Advances in Neural Information Processing Systems},
  34:14967--14979, 2021.

\bibitem{le2020contrastive}
Phuc~H Le-Khac, Graham Healy, and Alan~F Smeaton.
\newblock Contrastive representation learning: A framework and review.
\newblock {\em IEEE Access}, 2020.

\bibitem{LEI2022111141}
Maggie Lei, Hao Li, Ji~Li, Namrata Aundhkar, and Dae-Kyoo Kim.
\newblock Deep learning application on code clone detection: A review of
  current knowledge.
\newblock {\em Journal of Systems and Software}, 184:111141, 2022.

\bibitem{li2021concise}
Xiaoxu Li, Zhuo Sun, Jing-Hao Xue, and Zhanyu Ma.
\newblock A concise review of recent few-shot meta-learning methods.
\newblock {\em Neurocomputing}, 456:463--468, 2021.

\bibitem{li2021semi}
Yue Li and Jiong Zhang.
\newblock Semi-supervised meta-learning for cross-domain few-shot intent
  classification.
\newblock In {\em Proceedings of the 1st Workshop on Meta Learning and Its
  Applications to Natural Language Processing}, pages 67--75, 2021.

\bibitem{liu2021can}
Chenyao Liu, Zeqi Lin, Jian-Guang Lou, Lijie Wen, and Dongmei Zhang.
\newblock Can neural clone detection generalize to unseen functionalitiesƒ.
\newblock In {\em 2021 36th IEEE/ACM International Conference on Automated
  Software Engineering (ASE)}, pages 617--629. IEEE, 2021.

\bibitem{liu2019roberta}
Yinhan Liu, Myle Ott, Naman Goyal, Jingfei Du, Mandar Joshi, Danqi Chen, Omer
  Levy, Mike Lewis, Luke Zettlemoyer, and Veselin Stoyanov.
\newblock Roberta: A robustly optimized bert pretraining approach.
\newblock {\em arXiv preprint arXiv:1907.11692}, 2019.

\bibitem{lu2021codexglue}
Shuai Lu, Daya Guo, Shuo Ren, Junjie Huang, Alexey Svyatkovskiy, Ambrosio
  Blanco, Colin Clement, Dawn Drain, Daxin Jiang, Duyu Tang, et~al.
\newblock Codexglue: A machine learning benchmark dataset for code
  understanding and generation.
\newblock {\em arXiv preprint arXiv:2102.04664}, 2021.

\bibitem{mou2016convolutional}
Lili Mou, Ge~Li, Lu~Zhang, Tao Wang, and Zhi Jin.
\newblock Convolutional neural networks over tree structures for programming
  language processing.
\newblock In {\em Thirtieth AAAI Conference on Artificial Intelligence}, 2016.

\bibitem{nafi2019clcdsa}
Kawser~Wazed Nafi, Tonny~Shekha Kar, Banani Roy, Chanchal~K Roy, and Kevin~A
  Schneider.
\newblock Clcdsa: cross language code clone detection using syntactical
  features and api documentation.
\newblock In {\em 2019 34th IEEE/ACM International Conference on Automated
  Software Engineering (ASE)}, pages 1026--1037. IEEE, 2019.

\bibitem{oord2018representation}
Aaron van~den Oord, Yazhe Li, and Oriol Vinyals.
\newblock Representation learning with contrastive predictive coding.
\newblock {\em arXiv preprint arXiv:1807.03748}, 2018.

\bibitem{pearce2022examining}
Hammond Pearce, Benjamin Tan, Baleegh Ahmad, Ramesh Karri, and Brendan
  Dolan-Gavitt.
\newblock Examining zero-shot vulnerability repair with large language models.
\newblock In {\em 2023 IEEE Symposium on Security and Privacy (SP)}, pages
  1--18. IEEE Computer Society, 2022.

\bibitem{perez2019cross}
Daniel Perez and Shigeru Chiba.
\newblock Cross-language clone detection by learning over abstract syntax
  trees.
\newblock In {\em 2019 IEEE/ACM 16th International Conference on Mining
  Software Repositories (MSR)}, pages 518--528. IEEE, 2019.

\bibitem{NEURIPS_DATASETS_AND_BENCHMARKS2021_a5bfc9e0}
Ruchir Puri, David Kung, Geert Janssen, Wei Zhang, Giacomo Domeniconi, Vladimir
  Zolotov, Julian~T Dolby, Jie Chen, Mihir Choudhury, Lindsey Decker, Veronika
  Thost, Veronika Thost, Luca Buratti, Saurabh Pujar, Shyam Ramji, Ulrich
  Finkler, Susan Malaika, and Frederick Reiss.
\newblock Codenet: A large-scale ai for code dataset for learning a diversity
  of coding tasks.
\newblock In J.~Vanschoren and S.~Yeung, editors, {\em Proceedings of the
  Neural Information Processing Systems Track on Datasets and Benchmarks},
  volume~1, 2021.

\bibitem{quiring2019misleading}
Erwin Quiring, Alwin Maier, Konrad Rieck, et~al.
\newblock Misleading authorship attribution of source code using adversarial
  learning.
\newblock In {\em USENIX Security Symposium}, pages 479--496, 2019.

\bibitem{rabin2021generalizability}
Md~Rafiqul~Islam Rabin, Nghi~DQ Bui, Ke~Wang, Yijun Yu, Lingxiao Jiang, and
  Mohammad~Amin Alipour.
\newblock On the generalizability of neural program models with respect to
  semantic-preserving program transformations.
\newblock {\em Information and Software Technology}, 135:106552, 2021.

\bibitem{reimers-gurevych-2019-sentence}
Nils Reimers and Iryna Gurevych.
\newblock Sentence-{BERT}: Sentence embeddings using {S}iamese {BERT}-networks.
\newblock In {\em Proceedings of the 2019 Conference on Empirical Methods in
  Natural Language Processing and the 9th International Joint Conference on
  Natural Language Processing (EMNLP-IJCNLP)}, pages 3982--3992, Hong Kong,
  China, November 2019. Association for Computational Linguistics.

\bibitem{roy2018benchmarks}
Chanchal~K Roy and James~R Cordy.
\newblock Benchmarks for software clone detection: A ten-year retrospective.
\newblock In {\em 2018 IEEE 25th International Conference on Software Analysis,
  Evolution and Reengineering (SANER)}, pages 26--37. IEEE, 2018.

\bibitem{shobha2021code}
G~Shobha, Ajay Rana, Vineet Kansal, and Sarvesh Tanwar.
\newblock Code clone detection—a systematic review.
\newblock {\em Emerging Technologies in Data Mining and Information Security},
  pages 645--655, 2021.

\bibitem{snell2017prototypical}
Jake Snell, Kevin Swersky, and Richard Zemel.
\newblock Prototypical networks for few-shot learning.
\newblock {\em Advances in neural information processing systems}, 30, 2017.

\bibitem{song2022comprehensive}
Yisheng Song, Ting Wang, Puyu Cai, Subrota~K Mondal, and Jyoti~Prakash Sahoo.
\newblock A comprehensive survey of few-shot learning: Evolution, applications,
  challenges, and opportunities.
\newblock {\em ACM Computing Surveys}, 2022.

\bibitem{sonnekalb2022generalizability}
Tim Sonnekalb, Bernd Gruner, Clemens-Alexander Brust, and Patrick M{\"a}der.
\newblock Generalizability of code clone detection on codebert.
\newblock In {\em Proceedings of the 37th IEEE/ACM International Conference on
  Automated Software Engineering}, pages 1--3, 2022.

\bibitem{svajlenko2014towards}
Jeffrey Svajlenko, Judith~F Islam, Iman Keivanloo, Chanchal~K Roy, and
  Mohammad~Mamun Mia.
\newblock Towards a big data curated benchmark of inter-project code clones.
\newblock In {\em 2014 IEEE International Conference on Software Maintenance
  and Evolution}, pages 476--480. IEEE, 2014.

\bibitem{svajlenko2016bigcloneeval}
Jeffrey Svajlenko and Chanchal~K Roy.
\newblock Bigcloneeval: A clone detection tool evaluation framework with
  bigclonebench.
\newblock In {\em 2016 IEEE international conference on software maintenance
  and evolution (ICSME)}, pages 596--600. IEEE, 2016.

\bibitem{svajlenko2021bigclonebench}
Jeffrey Svajlenko and Chanchal~K Roy.
\newblock Bigclonebench.
\newblock In {\em Code Clone Analysis}, pages 93--105. Springer, 2021.

\bibitem{tao2022c4}
Chenning Tao, Qi~Zhan, Xing Hu, and Xin Xia.
\newblock C4: Contrastive cross-language code clone detection.
\newblock In {\em Proceedings of the 30th IEEE/ACM International Conference on
  Program Comprehension}, pages 413--424, 2022.

\bibitem{vinyals2016matching}
Oriol Vinyals, Charles Blundell, Timothy Lillicrap, Daan Wierstra, et~al.
\newblock Matching networks for one shot learning.
\newblock {\em Advances in neural information processing systems},
  29:3630--3638, 2016.

\bibitem{walsh2022automated}
Reece Walsh, Mohamed~H Abdelpakey, Mohamed~S Shehata, and Mostafa~M Mohamed.
\newblock Automated human cell classification in sparse datasets using few-shot
  learning.
\newblock {\em Scientific Reports}, 12(1):2924, 2022.

\bibitem{wang2022bridging}
Deze Wang, Zhouyang Jia, Shanshan Li, Yue Yu, Yun Xiong, Wei Dong, and Xiangke
  Liao.
\newblock Bridging pre-trained models and downstream tasks for source code
  understanding.
\newblock In {\em Proceedings of the 44th International Conference on Software
  Engineering}, pages 287--298, 2022.

\bibitem{wang2021grad2task}
Jixuan Wang, Kuan-Chieh Wang, Frank Rudzicz, and Michael Brudno.
\newblock Grad2task: Improved few-shot text classification using gradients for
  task representation.
\newblock {\em Advances in Neural Information Processing Systems},
  34:6542--6554, 2021.

\bibitem{wang2020detecting}
Wenhan Wang, Ge~Li, Bo~Ma, Xin Xia, and Zhi Jin.
\newblock Detecting code clones with graph neural network and flow-augmented
  abstract syntax tree.
\newblock In {\em 2020 IEEE 27th International Conference on Software Analysis,
  Evolution and Reengineering (SANER)}, pages 261--271. IEEE, 2020.

\bibitem{wang2021syncobert}
Xin Wang, Yasheng Wang, Fei Mi, Pingyi Zhou, Yao Wan, Xiao Liu, Li~Li, Hao Wu,
  Jin Liu, and Xin Jiang.
\newblock Syncobert: Syntax-guided multi-modal contrastive pre-training for
  code representation.
\newblock {\em arXiv preprint arXiv:2108.04556}, 2021.

\bibitem{wang-etal-2021-codet5}
Yue Wang, Weishi Wang, Shafiq Joty, and Steven~C.H. Hoi.
\newblock {C}ode{T}5: Identifier-aware unified pre-trained encoder-decoder
  models for code understanding and generation.
\newblock In {\em Proceedings of the 2021 Conference on Empirical Methods in
  Natural Language Processing}, pages 8696--8708, Online and Punta Cana,
  Dominican Republic, November 2021. Association for Computational Linguistics.

\bibitem{wei2017supervised}
Huihui Wei and Ming Li.
\newblock Supervised deep features for software functional clone detection by
  exploiting lexical and syntactical information in source code.
\newblock In {\em IJCAI}, pages 3034--3040, 2017.

\bibitem{wi2022hiddencpg}
Seongil Wi, Sijae Woo, Joyce~Jiyoung Whang, and Sooel Son.
\newblock Hiddencpg: large-scale vulnerable clone detection using subgraph
  isomorphism of code property graphs.
\newblock In {\em Proceedings of the ACM Web Conference 2022}, pages 755--766,
  2022.

\bibitem{xue2022seed}
Zhipeng Xue, Zhijie Jiang, Chenlin Huang, Rulin Xu, Xiangbing Huang, and Liumin
  Hu.
\newblock Seed: Semantic graph based deep detection for type-4 clone.
\newblock In {\em International Conference on Software and Software Reuse},
  pages 120--137. Springer, 2022.

\bibitem{ye2020misim}
Fangke Ye, Shengtian Zhou, Anand Venkat, Ryan Marcus, Nesime Tatbul,
  Jesmin~Jahan Tithi, Niranjan Hasabnis, Paul Petersen, Timothy Mattson, Tim
  Kraska, et~al.
\newblock Misim: A neural code semantics similarity system using the
  context-aware semantics structure.
\newblock {\em arXiv preprint arXiv:2006.05265}, 2020.

\bibitem{ye2021train}
Han-Jia Ye and Wei-Lun Chao.
\newblock How to train your maml to excel in few-shot classification.
\newblock {\em arXiv preprint arXiv:2106.16245}, 2021.

\bibitem{yefet2020adversarial}
Noam Yefet, Uri Alon, and Eran Yahav.
\newblock Adversarial examples for models of code.
\newblock {\em Proceedings of the ACM on Programming Languages},
  4(OOPSLA):1--30, 2020.

\bibitem{yu2023graph}
Dongjin Yu, Quanxin Yang, Xin Chen, Jie Chen, and Yihang Xu.
\newblock Graph-based code semantics learning for efficient semantic code clone
  detection.
\newblock {\em Information and Software Technology}, 156:107130, 2023.

\bibitem{yu2019neural}
Hao Yu, Wing Lam, Long Chen, Ge~Li, Tao Xie, and Qianxiang Wang.
\newblock Neural detection of semantic code clones via tree-based convolution.
\newblock In {\em 2019 IEEE/ACM 27th International Conference on Program
  Comprehension (ICPC)}, pages 70--80. IEEE, 2019.

\bibitem{yuan2022java}
Dawei Yuan, Sen Fang, Tao Zhang, Zhou Xu, and Xiapu Luo.
\newblock Java code clone detection by exploiting semantic and syntax
  information from intermediate code-based graph.
\newblock {\em IEEE Transactions on Reliability}, 2022.

\bibitem{zhang2023efficient}
Aiping Zhang, Liming Fang, Chunpeng Ge, Piji Li, and Zhe Liu.
\newblock Efficient transformer with code token learner for code clone
  detection.
\newblock {\em Journal of Systems and Software}, 197:111557, 2023.

\bibitem{zhang2020generating}
Huangzhao Zhang, Zhuo Li, Ge~Li, Lei Ma, Yang Liu, and Zhi Jin.
\newblock Generating adversarial examples for holding robustness of source code
  processing models.
\newblock In {\em Proceedings of the AAAI Conference on Artificial
  Intelligence}, volume~34, pages 1169--1176, 2020.

\bibitem{zhang2019novel}
Jian Zhang, Xu~Wang, Hongyu Zhang, Hailong Sun, Kaixuan Wang, and Xudong Liu.
\newblock A novel neural source code representation based on abstract syntax
  tree.
\newblock In {\em 2019 IEEE/ACM 41st International Conference on Software
  Engineering (ICSE)}, pages 783--794. IEEE, 2019.

\bibitem{zhang2022graph}
Ruiheng Zhang, Shuo Yang, Qi~Zhang, Lixin Xu, Yang He, and Fan Zhang.
\newblock Graph-based few-shot learning with transformed feature propagation
  and optimal class allocation.
\newblock {\em Neurocomputing}, 470:247--256, 2022.

\bibitem{zhang2023challenging}
Weiwei Zhang, Shengjian Guo, Hongyu Zhang, Yulei Sui, Yinxing Xue, and Yun Xu.
\newblock Challenging machine learning-based clone detectors via
  semantic-preserving code transformations.
\newblock {\em IEEE Transactions on Software Engineering}, 2023.

\end{thebibliography}

\end{document}